\documentclass{ws-procs9x6}

\begin{document}
%
%
\def\cal{\mathcal}
\def\half{\scriptstyle{\frac{1}{2}}}
\def\halft{\textstyle{\frac{1}{2}}}
\def\osqrt{\textstyle{\frac{1}{\sqrt2}}}
\def\lsqrt{\textstyle{\frac{\l}{\sqrt2}}}
\def\phalf{\textstyle{\frac{\pi}{2}}}
\newcommand{\fscr}[2]{\scriptstyle \frac{#1}{#2}}
\def\cc{\hbox{C\kern-0.55em\raise0.4ex\hbox{$\scriptstyle |$}}}
\def\zz{\hbox{{\sf Z}\kern-0.45em\raise0.0ex\hbox{\sf Z}}}
\def\nat{\hbox{{\sf |}\kern-0.45em\raise0.0ex\hbox{\sf Í}}}
\def\rr{\hbox{R\kern-0.55em\raise0.4ex\hbox{$\scriptstyle |$}}}
\def\ena{1\hskip-0.25truecm 1}
\def\zita{{\zz}^{ 2}_{N}}
\def\zitas{{\zz}^{* 2}_{N}}
\def\zi{{\zz}^{2r+1}_{2}}
\def\zist{{\zz}^{* 2r+1}_{2}}
\def\ccd{{\cc\hskip0.2truecm}^{2}}
\def\cct{{\cc\hskip0.2truecm}^{3}}
\def\en{{\bf 1 \/}}
\def\pf{{\it pd \/}}
\def\ie{{\it i.e \/}}
\def\tr{{\rm Tr \/}}
\def\eg{{\it e.g \/}}
\def\cf{{\it c.f \/}}
\def\viz{{\it viz. \/}}
\def\aa {{\cal A \/}}
\def\ad{a^\dag}
\def\ab{\bar{\alpha}}
\def\fh{\hat{f}}
\def\nub{\bar{\nu}}
\def\hi{\chi_{klm}}
\def\udp{U_{\lambda}^{\dagger}}
\def\udm{U_{-\lambda}^{\dagger}}
\def\utp{\tilde{U}_{\lambda}}
\def\utm{\tilde{U}_{-\lambda}}
\def\up{U_{\lambda}}
\def\um{U_{-\lambda}}
\def\td{A^\dag_q}
\def\so{\sigma_{1}}
\def\st{\sigma_{2}}
\def\sth{\sigma_{3}}
\def\sp{\sigma^{+}}
\def\sm{\sigma^{-}}
\def\ovl{\overline}

%
\newcommand{\bra}[1]{\left<#1\right|} \newcommand{\ket}[1]{\left|#1\right>}
\newcommand{\braket}[1]{\left<#1\right>}
\newcommand{\inner}[2]{\left<#1|#2\right>}
\newcommand{\sand}[3]{\left<#1|#2|#3\right>}
\newcommand{\proj}[2]{\left|#1\left>\right<#2\right|} %
\newcommand{\rbra}[1]{\left(#1\right|} \newcommand{\rket}[1]{\left|#1\right)}
\newcommand{\rbraket}[1]{\left(#1\right)}
\newcommand{\rinner}[2]{\left(#1|#2\right)}
\newcommand{\rsand}[3]{\left(#1|#2|#3\right)}
\newcommand{\rproj}[2]{\left|#1\left)\right(#2\right|}
\newcommand{\absqr}[1]{{\left|#1\right|}^2}
\newcommand{\abs}[1]{\left|#1\right|}
\newcommand{\pl}[2]{\partial_{#1}^{#2}}
\newcommand{\plz}[1]{\partial_{z}^{#1}}
\newcommand{\plzb}[1]{\partial_{\overline{z}}^{#1}}
\newcommand{\zib}[1]{{\overline{z}}^{#1}}
\newcommand{\mat}[4]{\left(\begin{array}{cc} #1 & #2 \\ #3 & #4
\end{array}\right)} %
\newcommand{\col}[2]{\left( \begin{array}{c} #1 \\ #2
\end{array} \right)}
\newcommand{\qcol}[2]{\left[ \begin{array}{c} #1 \\ #2
\end{array} \right]_q }
\def\a{\alpha}
\def\b{\beta}
\def\g{\gamma}
\def\d{\delta}
\def\e{\epsilon}
\def\z{\zeta}
\def\th{\theta}
\def\f{\phi}
\def\la{\lambda}
\def\m{\mu}
\def\p{\pi}
\def\om{\omega}
\def\D{\Delta}
\def\zb{\bar{z}}
\newcommand{\lag}[2]{L_{#1}^{#2}(4\l^{2})}
\newcommand{\mes}[1]{d\mu(#1)}
\def\nd{\noindent}
\def\nn{\nonumber}
\def\cap{\caption}
\def\cline{\centerline}
\newcommand{\be}{\begin{equation}}
\newcommand{\ee}{\end{equation}}
\newcommand{\ba}{\begin{array}}
\newcommand{\ea}{\end{array}}
\newcommand{\bea}{\begin{eqnarray}}
\newcommand{\eea}{\end{eqnarray}}
\newcommand{\beann}{\begin{eqnarray*}}
\newcommand{\eeann}{\end{eqnarray*}}
\newcommand{\bfg}{\begin{figure}}
\newcommand{\efg}{\end{figure}}
\def\ucn{U_{\scriptstyle  CN}}
\def\uca{U_{\scriptstyle  f}}


\title{On Algebraic and Quantum Random Walks
\footnote{\uppercase{Q}uantum \uppercase{P}robability and
\uppercase{I}nfinite \uppercase{D}imensional \uppercase{A}nalysis:
 \uppercase{F}rom \uppercase{F}oundations to \uppercase{A}pplications, \uppercase{QP-PQ}
\uppercase{V}ol.18, eds. \uppercase{M.} \uppercase{S}ch{\"u}rmann
and \uppercase{U.} \uppercase{F}ranz, (\uppercase{W}orld
\uppercase{S}cientific, 2005), p. 174-200.}}

\author{Demosthenes Ellinas
}

\address{Technical University of Crete\\
Department of Sciences, Division of Mathematics, \\
GR-731 00 Chania Crete Greece\\
E-mail: ellinas@science.tuc.gr}



\maketitle

\abstracts{ Algebraic random walks (ARW) and quantum mechanical
random walks (QRW) are investigated and related. Based on minimal
data provided by the underlying bialgebras of functions defined on
e. g the real line \textbf{R}, the abelian finite group
$\textbf{Z}_N$, and the canonical Heisenberg-Weyl algebra
\textbf{hw}, and by introducing appropriate functionals on those
algebras, examples of ARWs are constructed. These walks involve
short and long range transition probabilities as in the case of
\textbf{R} walk, bistochastic matrices as for the case of
$\textbf{Z}_N$ walk, or coherent state vectors as in the case of
\textbf{hw} walk. The increase of classical entropy due to
majorization order of those ARWs is shown, and further their
corresponding evolution equations are obtained. Especially for the
case of \textbf{hw} ARW, the diffusion limit of evolution equation
leads to a quantum master equation for the density matrix of a
boson system interacting with a bath of quantum oscillators
prepared in squeezed vacuum state. A number of generalizations to
other types of ARWs and some open problems are also stated. Next,
QRWs are briefly presented together with some of their distinctive
properties, such as their enhanced diffusion rates, and their
behavior in respect to the relation of majorization to quantum
entropy. Finally, the relation of ARWs to QRWs is investigated in
terms of the theorem of unitary extension of completely positive
trace preserving (CPTP) evolution maps by means of auxiliary
vector spaces. It is applied to extend the CPTP step evolution map
of a ARW for a quantum walker system into a unitary step evolution
map for an associated QRW of a walker+quantum coin system.
Examples and extensions are provided.}

\newpage \section{Introduction} Random walks formulated in an algebraic
framework\cite{meyer,schurmann,majidbook,fs} of finite groups,
bialgebras and operator algebras as well as in the framework of
Quantum Mechanics\cite{ambainis}$-$\cite{bet}, and references
therein), are investigated. Minimal data for such constructions
consist of a bialgebra\cite{abe} and an integral (functional)
defined on it, or alternatively of some Lie algebra, and two
quantum systems modelling the walker and the coin system, together
with a map modelling the coin tossing, that decides
probabilistically  the stepping of the walker.

Examples of ARWs treated in the following subsections are walks on
algebras of functions on $\textbf{R}$, on ${\textbf{Z}}_{N}$, and
on the canonical algebra Heisenberg-Weyl
${\textbf{hw}}$\cite{appell}(section 2). For those walks we show
how to define the entropy functional of their respective integral
and/or Markov transition operator, and how to deduce that these
are entropy increasing random walks by using arguments based on
the interrelations relations between  majorization bistochastic
matrices and
entropy\cite{marshalolkin,bhatia,albertiuhlmann,nielsennotes}.
Moreover, a number theoretic decomposing of ARW on
${\textbf{Z}}_{N}$ is analysed, that is called prime
decomposition\cite{ellinasf}, and refers to its factorization into
products of similar smaller ${\textbf{Z}}_{N}$-walks.
Mathematically this decomposition is based on the Chinese
Remainder Theorem and the co-associativity property.  Also for the
ARWs on $\textbf{R}$, in addition to the usual case of short range
walk with nearest neighbor (NN) transitions (Polya
walk)\cite{hughes}, we discuss in our algebraic framework the
cases of: i) the NN centrally biased random walk (Gillis
walk)\cite{gillis} and
the case of symmetric random walk with exponentially distributed
steps (Linderberg-Shuler:LS walk)\cite{ls}. Finally, for the
${\textbf{hw}}$ ARW, where its functional is constructed by means
of the eigenstates of the annihilation operator of that algebra
i.e the family of coherent state vectors\cite{klauder,perelomov},
the continuous time, or diffusion like limit, is
obtained\cite{appell}. This limit results into a trace preserving
quantum master equation\cite{davies,lind} for the density matrix
of a quantum boson system, which physically is identified with the
evolution equation of an open quantum boson system interacting
coherently with a classical electric field and incoherently
(dissipative interaction) with a bath of quantum oscillators
\textit{rigged} initially into a squeezed vacuum state (squeezed
white noise)\cite{scully,stenholm,gea}.

Section 3, gives a concise prescription of the concept of QRW,
using the example of QRW on integers as paradigm\cite{bet}. It
briefly explains the notion of quantum coin system and the coin
tossing map, and summarizes two emblematic properties of that
walk, namely the quadratic enhancement of its diffusion rate due
to quantum entanglement between the walker and coin systems, and
the entropy increase without majorization effect of its
probability distributions (pd). This section ends with a group
theoretical scheme of classification of various known QRWs.

In section 4, a relation connecting ARW and QRW is put forward.
The connection is grounded on the theorem due to Naimark that
asserts the possibility of implementing in a unitary way a CPTP
map operating on e.g the density operator of a quantum
system\cite{kraus}. This unitary extension is realized in the
original vector space of the density operator augmented by an
auxiliary vector space, the ancilla space in the terminology of
Quantum Information theory\cite{nielsen}. Applied in the context
of CPTP of a ARW, the ancilla space is identified with the quantum
coin state space of a QRW. The Kraus generators determining the
CPTP map of a ARW serve to built, albeit in non unique way,  the
unitary evolution operator of the associated QRW. This section
concludes with the example of an explicit construction of a QRW
associated to the ${\textbf{hw}}$ ARW, of section 2.

Finally, section 4, summarized some of the results and gives some
prospected applications of the ARW-QRW concepts and formalism.

\section{Algebraic Random Walks}
\subsection{The case of {\bf R}}

\textit{Proposition 1}. Let the bialgebra of real formal power
series $H\equiv Fun(\mathbf{R})$ generated by the coordinate
function $X,$ and let the positive definite functional
$\phi:H\rightarrow\mathbf{R}$, defined as $\phi=\sum_{i\in
Z}p_{i}\phi_{\alpha_{i}},$ with $0\leq p_{i}\leq1,\sum_{i\in
Z}p_{i}=1,$ where $\phi_{\alpha_{i}}(f(x))=f(\alpha_{i}),$ the
functional that evaluates any function $f\in H,$ at the point
$\alpha_{i}=i\alpha,$ defined for some step $\alpha\in$
$\mathbf{R}_{+},$ for $i\in Z.$ The $n$-step convoluted functional
becomes \be \phi^{\ast n}=\sum_{i\in Z}p_{i}^{(n)}
\phi_{\alpha_{i}}=\widetilde{\phi}^{T}D^{n-1}p, \ee \nd where
$p^{(n)}=D^{n-1} p^{(1)},$ with the stochastic column vectors
$p^{(k)}=(p_{i}^{(k)})_{i\in Z},k=1,2,3,\ldots$ and initially
$p^{(1)}\equiv p=(p_{i})_{i\in Z}$. Also $D=(D_{ij})=(p_{i-j}),$
$i,j\in Z$ a bistochastic infinite matrix (delta matrix), and
$\widetilde{\phi}=(\phi_{\alpha_{i}})_{i\in Z}$ a column vector.
Majorization ordering among pd's is valid at each step i.e
$p^{(n+1)}\prec p^{(n)},$ and consequently the ARW is entropy
increasing, namely, $S(\phi^{\ast n+1})\geq S(\phi^{\ast n}),$
$S(T_{\phi}^{n+1})\geq S(T_{\phi}^{n}),$ where $S(\phi^{\ast
n})\equiv S(T_{\phi}^{n})\equiv$ $S(p^{(n)}),$ with $S(x)$ any
Shur-convex function e.g the classical entropy i.e
$S(x)=-\sum_{i}x_{i}\log x_{i}.$

\textit{\bigskip Proof: }Operating with convoluted functionals on
some function $f\in
H,$(\cite{majid1,majid2},\cite{qnoice1,qnoice2,qnoice3,qnoice4},\cite{elltso1,elltso2})
yields
\begin{align}
\phi^{\ast2}  &  =(\phi\otimes\phi)\circ\Delta=\sum_{i,j\in Z}p_{i}^{(1)}%
p_{j}^{(1)}\phi_{\alpha_{i}+\alpha_{j}}\nonumber\\
&  =\sum_{k,j\in
Z}p_{i}^{(1)}p_{k-i}^{(1)}\phi_{\alpha_{k}}\equiv\sum_{k,j\in
Z}p_{i}^{(1)}D_{ki}\phi_{\alpha_{k}}=\widetilde{\phi}^{T}Dp.
\end{align}

\bigskip By induction we obtain the aimed relation $\phi^{\ast n}%
=\widetilde{\phi}^{T}p^{(n)}=\widetilde{\phi}^{T}D^{n-1}p^{(1)}.$
Two important properties of the delta matrix are: i)
\textit{bistochasticity}, i.e the column and row sums is one,
which is expressed by means of the column vector of units
$e=(\ldots,1,1,1,\ldots),$ that is left and right eigenvector of
$D,$ i.e $De=e,$ $e^{T}=e^{T}D,$ and ii) the \textit{shift
property }i.e $D_{i,j}=p_{i-j}=p_{i+1-(j+1)}=D_{i+1,j+1}.$ This
property in the case of ARW in $Z$ governed by a pd with finite
support, or in the case of a $N\subset Z$ finite dimensional walk
(see remark below), amounts to a  bistochastic matrix $D.$

The functional of the walk at each time step $\phi^{\ast n}=\sum_{i\in Z}%
p_{i}^{(n)}\phi_{\alpha_{i}},$ is characterized by the pd $p^{(n)}%
=(p_{i}^{(n)})_{i\in Z}$ , which in turn is determined by the
bistochastic matrix $D$ i.e $\ p^{(n)}=Dp^{(n-1)}=D^{n-1}p^{(1)}.$
Let us assume that the pd's are of finite support (but see remark
below), then by invocation of the theorem stating that two
discrete pd's $x=(x_{m})_{m},$ $y=(y_{k})_{k},$ that are connected
by a bistochastic matrix $D$ i.e $x=Dy,$ are ordered by
majorization
\cite{marshalolkin,bhatia,albertiuhlmann,nielsennotes} i.e $x$
$\prec y$, we conclude that the sequence of pd's
$\{p^{(1)},p^{(2)},p^{(3)},\ldots\}$ , of site occupation
probabilities resulting at each time step during the evolution of
the walk is partially ordered by majorization i. e $p^{(n)}\prec
p^{(n-1)},$ $n=1,2,3,\ldots$ .

Let us adopt now the definition of the entropy of a functional to
be the entropy of the pd that determines that functional i.e we
set $S(\phi^{\ast n})\equiv S(T_{\phi}^{n})\equiv$ $S(p^{(n)}),$
or more generally we do so for any convex function
$S:\mathbf{R}\rightarrow\mathbf{R}$ of the type of the so called
Shur-convex functions e.g the classical Shannon entropy $S(p)=\sum
_{i}p_{i}\log p_{i}$, or the functions $F(p)=\sum_{i}p_{i}^k$, for
any constant $k\leq 1$, or $F(p)=-\prod_{i}p_{i}$\cite{bhatia}.

By virtue of the theorem stating that $x$ $\prec y$ implies
$F(x)<F(y)$ , where $F(x)=\sum_{i}f(x_{i}),$ for any convex
function $f:\mathbf{R}\rightarrow\mathbf{R,}$ or otherwise said
that the convex functions \textit{isotonic} to
majorization\cite{bhatia}, we conclude that for the pd resulting
from the random walk the majorization ordering is valid at each
step i.e. $p^{(n)}\prec p^{(n-1)},$ $n=1,2,3,\ldots$ , and this
implies ordering for e.g their entropies i.e.
$S(p^{n})>S(p^{n-1}),$ and similarly for their functionals and
transition operators. As majorization order implies entropy
increase, it is considered as a measure of disorder, and this
allow us to conclude the ARW are getting more disordered in the
course of time with respect to their site-visiting pd's, which are
getting more entropic, approaching, if left uninterrupted, to the
uniform distribution of maximal entropy. \ \ \ \ \ \ \ \ \ \ \ \ \
\ \ \ \ \ \ \ \ \ \ \ \ \ \ \ \ \ \ \ \ \ \ \ \ \ \ \ \ \ \ \ \ \
\ \ \ \ \ \ \ \ \ \ \ \ \ \ \ \ \ \ \ \ \ \ \ \ \ \ \ \ \ \ \ \ \
\ \ \ \ \ \ \ \ \ \ \ \ \ \ \ \ \ \ \ \ \ \ \ \ \ \ \ \ \ $\square
$\qquad
$\ \ \ \ \ \ \ \ \ \ \ \ \ \ \ \ \ \ \ \ \ \ \ \ \ \ \ \ \ \ \ \ \ \ \ \ \ \ \ \ \ \ \ \ \ \ \ \ \ \ \ \ \ \ \ \ \ \ \ \ $%

\bigskip Remarks: 1) The assumption in the previous proof about the support of
the pd's been finite is not actually necessary. In fact in the
proof given by Hardy et. al \cite{hardy} \ is stated that given
$p\prec q$ for finite sequence of $q,p$ pd's, use of Muirhead's
algorithm leads to a bistochastic matrix $A,$ such that $q=Ap,$
and then the proposition: $H(Ap)\geq H(p)$ for $\ H$
information/entropy or more generally a Shur convex function, is
applied. This proof is constructive and builds $A$ in a number of
steps not greater than the lengths of $p,$ $q$, therefore for
infinite pd's a theorem not using bistochastic operators for the
characterization of majorization is needed. Such a theorem is
provided in \cite{sikic} .

2) For the above proposition the corresponding Markov transition
operator defined as $T_{\phi}=(\phi$ $\otimes id)\circ\Delta,$ is
equal to $T_{\phi }=\sum_{i\in Z}p_{i}e^{\alpha_{i}\frac{d}{dx}}.$
For the simplest case of $p_{\pm}=(p,1-p),$
$\alpha_{\pm}=\pm\alpha,$ and all other $p$'s and
$\alpha^{\prime}$ s been zero, the continues limit
$lim_{n\rightarrow\infty }=(T_{\phi})^{n}\equiv T_{\phi}^{\infty}$
\ has been obtained that leads to a diffusion
equation\cite{majid1,majidbook,fs,appell} .

3) The above general algebraic setting implies that the stepping
probability matrix $D=(D_{ij})=(p_{i-j}),$ $i,j\in\mathbf{Z,}$
would be expanded in the enveloping algebra
$\mathcal{U}(\mathbf{e}(2))$ of the Euclidean Lie algebra
$\mathbf{e}(2)\thickapprox\mathbf{iso}(2)$, spanned by monomials
of its generators $\{E_{+},E_{-},L\},$ that satisfy the defining
commutation relations \cite{gilmore}
\begin{equation}
\lbrack E_{+},E_{-}]=0,\qquad\lbrack L,E_{\pm}]=\pm E_{\pm}.
\end{equation}

An irreducible matrix representation of those generators in the
Hilbert space $\mathcal{H}\equiv l_{2}(\mathbf{Z})$ spanned by the
eigenvectors of the "distance operator" $L,$ is useful in
expressing the $D$ matrix for various random walks, and look for
solutions by means of e.g the Fourier method. This irrep in the
canonical basis of $\mathcal{H},$ and using the same symbol for
abstract generators and their matrices reads:%
\begin{equation}
L=\sum_{m\in Z}me_{m}e_{m}^{\dagger},\text{ \ \ }E_{\pm}=\sum_{m\in Z}%
e_{m\pm1}e_{m}^{\dagger}.
\end{equation}

Next we present three different random walks in $\mathbf{R,}$ that
can be used as show cases of the scheme of ARW presented here. In
these concrete examples the defining the walk transition
probability matrix $D,$ is written as an element of the
$\mathcal{U}(\mathbf{e}(2))$ algebra. No attempt will be made to
give an algebraic solution for the problem of finding the $n$
th$-$step site occupancy probability distribution, as this can be
solved by other means.

The examples include: i) the simplest case of symmetric
nearest-neighbor (NN) random walk (Polya-walk\cite{hughes}); one
of its deformations, ii) the NN centrally (site $n=0)$ biased
random walk (Gillis-walk\cite{gillis}), which refers to a solvable
case of a walk with no translational invariance, and stepping
probabilities with a bias which has \ power law decay, or more
specifically which decays in proportion from the origin of
coordinates. The $\ \varepsilon-$deformation parameter is chosen
so that when $\varepsilon>0,$ the walk is biased to enhance
returns to the origin, while if $\varepsilon<0,$ escape from the
origin is enhanced, and iii) a symmetric random walk with non-
nearest-neighbor transitions with transition step length decaying
according to an exponential law (LS-walk\cite{ls}).
Explicitly we have:

1) \bigskip\textit{Polya-walk: }symmetric nearest-neighbor(NN)
random walk
with Markov transition operator%
\begin{equation}
D_{P}=\frac{1}{2}(E_{-}+E_{+})
\end{equation}
with matrix elements the inter-site transition probabilities%

\begin{equation}
D_{P}(l,l^{^{\prime}})=\frac{1}{2}\delta_{l,l^{^{\prime}}-1}+\frac{1}{2}%
\delta_{l,l^{^{\prime}}+1}.
\end{equation}

2) \textit{Gillis-walk: }nearest-neighbor centrally (site $n=0)$
biased random walk with Markov transition operator\textit{ }
\begin{equation}
D_{G}=D_{G}(\varepsilon)=\frac{1}{2}(E_{-}+E_{+})+(E_{-}-E_{+})\frac
{\varepsilon}{2N}P_{0}^{\bot},\text{ \ \ \ \ \ \ \ \ \
}-1<\varepsilon<1
\end{equation}
where $P_{0}=e_{0}e_{0}^{\dagger},$ $P_{0}^{\bot}=1-P_{0},$ the
projection operators in the vector $e_{0},$ and its orthogonal
subspace respectively in
$l_{2}(\mathbf{Z}),$ and with matrix elements the inter-site transition probabilities%

\begin{align}
D_{G}(l,l^{^{\prime}})  &  =\frac{1}{2}(1+\frac{\varepsilon}{l^{^{\prime}}%
})\delta_{l,l^{^{\prime}}-1}+\frac{1}{2}(1-\frac{\varepsilon}{l^{^{\prime}}%
})\delta_{l,l^{^{\prime}}+1},\text{ \ \ \ \ \ \ \ }l^{^{\prime}}\neq0\\
D_{G}(l,0)  &  =\frac{1}{2}\delta_{l,-1}+\frac{1}{2}\delta_{l,1}.
\end{align}



3) \textit{LS-walk: }symmetric random walk with exponentially
distributed
steps with Markov transition operator%
\begin{equation}
D_{LS}=D_{LS}(\varepsilon)=\frac{e^{\varepsilon}-1}{2}\sum_{k\in Z_{>0}%
}e^{-k\varepsilon}(E_{-}^{k}+E_{+}^{k})
\end{equation}
with matrix elements the inter-site transition probabilities%

\begin{align}
D_{LS}(l,l^{^{\prime}})  &
=\frac{e^{\varepsilon}-1}{2}e^{-\left\vert
l-l^{^{\prime}}\right\vert \varepsilon},\qquad\left\vert l-l^{^{\prime}%
}\right\vert >0\\
&  =0,\qquad\qquad\qquad\qquad\ l=l^{^{\prime}}.\nonumber
\end{align}

Use of the eigenvector equations $E_{\pm}e=e,$ $\
e^{\dagger}=e^{\dagger }E_{\pm},$ leads to the conclusion that the
transition matrices of the Polya, and LS random walks i.e the
matrices $D_{P},D_{R},$ and $D_{LS}$ respectively are
bistochastic, while that of the Gillis walk $D_{G},$ is column
stochastic. Also from the definitions the following two limits are
deduced:
$\lim_{\varepsilon\rightarrow0}D_{G}(\varepsilon)=\lim_{\varepsilon
\rightarrow\infty}D_{LS}(\varepsilon)=D_{P}.$\bigskip\ \ \ \ \ \ \

\textit{Generalizations:} The \textit{ }2D generalized Gillis
random walk with entanglement. This model describes a 2D NN random
walk with a biased towards a point placed at $(m_{1},m_{2})$
coordinates on the plane. The bias is an attractive or repelling
one depending on the sign of two parameters
$(\varepsilon_{1},\varepsilon_{2})$ \ in reference to the motion
along $x,y$ $\ $axes respectively, and its strength decays
following an inverse power law with characteristic exponents
($a_{1},a_{2})$ correspondingly. This is summarized by writing
explicitly all parameters in the Markov transition
operator\textit{ }$D_{G}^{2D}(\varepsilon_{1},$ \ $m_{1},$ $a_{1}%
;\varepsilon_{2},\ m_{2},a_{2}),$ and their domain of values
$\{\varepsilon _{j},$ \ $m_{j},$ $a_{j}\}\in\{$
$(-1,1),\mathbf{Z},\mathbf{Z}_{+}\},$ $j=1,2.$ The transition
matrix along each axis is
\begin{equation}
D_{G}(\varepsilon_{j},m_{j},a_{j})=\frac{1}{2}(E_{-}+E_{+})+(E_{-}-E_{+}%
)\frac{\varepsilon_{j}}{2(N-m_{j}\mathbf{1})^{a_{j}}}P_{j}^{\bot},\text{
\ \ \ \ }j=1,2
\end{equation}
where $P_{j}=e_{j}e_{j}^{\dagger},$ $P_{j}^{\bot}=1-P_{j},$ the
projection operators in the vector $e_{j},$ and its orthogonal
subspace respectively in $l_{2}(\mathbf{Z}).$ The 2D transition
operator consists of an entangled\cite{nielsen} convex combination
of two factorizable 1D transition operators \ with the parameters
of position of bias site, characteristic decay exponent, and decay
strength c.f $\ \{m_{j},a_{j},\varepsilon_{j},\},$ $j=1,2$ \
interchanged, it reads ($0\leq q\leq1)$

\begin{align}
D_{G}^{2D}  &  =qD_{G}(\varepsilon_{1},m_{1},a_{1})\otimes D_{G}%
(\varepsilon_{2},m_{2},a_{2})\nonumber\\
&  +(1-q)D_{G}(\varepsilon_{2},m_{2},a_{2})\otimes
D_{G}(\varepsilon_{1} ,m_{1},a_{1}),
\end{align}
with matrix elements the inter-site transition probabilities

\begin{align}
D_{G}^{2D}(l_{1},l_{1}^{^{\prime}};l_{2},l_{2}^{^{\prime}})  &  =q\frac{1}%
{4}\left[
1+\frac{\varepsilon_{1}}{(l_{1}^{^{\prime}}-m_{1})^{a_{1}}}\right]
\left[
1+\frac{\varepsilon_{2}}{(l_{2}^{^{\prime}}-m_{2})^{a_{2}}}\right]
\delta_{l_{1},l_{1}^{^{\prime}}-1}\delta_{l_{2},l_{2}^{^{\prime}}%
-1}\nonumber\\
&  +(1-q)\frac{1}{4}\left[  1+\frac{\varepsilon_{2}}{(l_{1}^{^{\prime}}%
-m_{2})^{a_{2}}}\right]  \left[  1+\frac{\varepsilon_{1}}{(l_{2}^{^{\prime}%
}-m_{1})^{a_{1}}}\right]  \delta_{l_{1},l_{1}^{^{\prime}}-1}\delta
_{l_{2},l_{2}^{^{\prime}}-1,}%
\end{align}
with $(l_{1}^{^{\prime}},l_{2}^{^{\prime}})\neq(m_{1},m_{2}),$ for
the $q$ \ component and
$(l_{1}^{^{\prime}},l_{2}^{^{\prime}})\neq(m_{2},m_{1}),$ for the
$(1-q)$ component of the convex combination. Also
\begin{align}
 D_{G}^{2D}(l_{1},l_{2}^{^{\prime}}   & =m_{1,2};l_{1},l_{2}^{^{\prime}%
}=m_{2,1})=q(\frac{1}{2}\delta_{l_{1},m_{1}-1}+\frac{1}{2}\delta_{l_{1}%
,m_{1}+1})\nn \\ . (\frac{1}{2}\delta_{l_{2},m_{2}-1}+\frac{1}{2}\delta_{l_{2},m_{2}%
+1})
&  +(1-q)(\frac{1}{2}\delta_{l_{1},m_{2}-1}+\frac{1}{2}\delta_{l_{1},m_{2}%
+1})\nn
\\ .(\frac{1}{2}\delta_{l_{2},m_{1}-1}+\frac{1}{2}\delta_{l_{2},m_{1}+1}).
\end{align}
the matrix elements with $(l_{1}^{^{\prime}},l_{2}^{^{\prime}})=(m_{1}%
,m_{2}),$ for the $q$ \ component and $(l_{1}^{^{\prime}},l_{2}^{^{\prime}%
})=(m_{2},m_{1}),$ for the $(1-q)$ component of the convex
combination. The role of bias and that of the entanglement of the
two 1D walks, can be investigated in an algebraic manner in terms
of tensor product representations of the $\mathbf{iso(2)}$
algebra, and it will be given elsewhere\cite{ellinasetal}.

\subsection{The case of ${\bf Z_N}$}
In this section we give a brief study of algebraic random walks on
abelian groups $\mathbf{Z}_{N}$ \cite{majidbook}, using their
underlying bialgebra structure, and further investigate possible
forms of their decomposition into simpler and dimensionally lower
ARWs, based on number theoretic properties of $N$.

\textit{Proposition 2}. Let the multiplicative abelian group $\mathbf{Z}%
_{N}=\{e,g,g^{2},\ldots,g^{N-1}\}$ and the bialgebra $H\equiv
\mathbf{C(\mathbf{Z}_{N})}\equiv
Fun(\mathbf{Z}_{N})=\mathbf{C}-span\langle \left\{  g^{i}\right\}
_{i=0}^{N-1}\rangle$, with dual algebra $H^{\ast
}\equiv\mathbf{CZ}_{N}\equiv Fun(Z_{N}^{\ast})$, \ with pairing
given by evaluation. Let the positive definite functional (state)
of a $\mathbf{Z}_{N}$ random walk
$\phi=\sum_{i\in\lbrack0,N-1]}p_{i}g^{i},$ identified as weighted
sum of elements of $H,$ with $0\leq p_{i}\leq1,\sum_{i\in\lbrack0,N-1]}%
p_{i}=1.$ By means of the column vectors $P=(p_{i})_{i\in\lbrack
0,N-1]},G=(g^{i})_{i\in\lbrack0,N-1]},$we write it as
$\phi=G^{T}P,$ where $T$ denotes transpose. Then the $n$-th step
convolution becomes $\phi^{\ast n}=G^{T}D^{n-1}P$, where $\ $\
$D=(D_{ij})=(p_{i-j\mid\operatorname{mod}N}),$
$i,j\in\lbrack0,N-1]$ a circulant bistochastic matrix, to be
called delta matrix.

\bigskip\textit{Proof: }Straightforward (c.f. \cite{majidbook}).

\bigskip\textit{Remarks}: 0) The delta matrix is more precisely
a circulant bistochastic matrix\cite{circ}, that can be treated as
an element of finite Heisenberg group$H_{N}$ \cite{ellinasf}, and
this leads to an explicit solution for the dynamics of
$\mathbf{Z}_{N}$ walk.

1) Recall the following version of the Chinese Remainder Theorem
(CRT)\cite{schroeder}: let $N=N_{1}N_{2}$, the decomposition of
positive integer $N,$ into product of coprimes $N_{1},N_{2},$ then
working with the abelian additive groups of numbers $Z_{N_{1,2}},$
$\operatorname{mod}N_{1,2}$ respectively, we can introduce the
unique bijection
\begin{equation}
\delta:Z_{N}\longrightarrow Z_{N_{1}}\times Z_{N_{2}},\text{ \ \ }%
\delta(x)=(x-\rho N_{1},x-\sigma N_{2}),
\end{equation}
$\rho,\sigma\in Z,$ that maps the numbers of $Z_{N},$ into the
ordered pair of its remainders after division by $N_{1},N_{2}.$
Its dual map is the inverse $\mu\equiv\delta^{-1}=Z_{N_{1}}\times
Z_{N_{2}}\longrightarrow Z_{N},$which constructs the unique number
$x$ from its remainders $(a,b)$, with respect to
the divisors $N_{1},N_{2},$ as $\mu(a,b)=aN_{1}^{\varphi(N_{1})}%
+bN_{2}^{\varphi(N_{2})}=x,$ where $\varphi(c)=\#\left\vert c\geq
m\in Z_{+};GCD(m,c)=1\right\vert ,$ the Euler function of given
integer $c,$ that equals the number of co-primes less or equal to
$c.$

2) The same factorization is valid for abelian multiplicative
groups i.e
$\mathbf{Z}_{N}\cong\mathbf{Z}_{N_{1}}\bigotimes\mathbf{Z}_{N_{2}},$
if $N=N_{1}N_{2},$ and $N_{1},N_{2},$ are relative primes.

3) Let for later use introduce now the map $\delta\longrightarrow
V_{\delta},$ that uniquely determines from the CRT bijection
$\delta,$ with $\delta
(i)=(i_{1},i_{2}),$ the isometric matrix $V_{\delta}:\mathbf{C}_{N}%
\longrightarrow\mathbf{C}_{N_{1}}\otimes\mathbf{C}_{N_{2}},$
written in the canonical basis as\cite{ellinasf}
\begin{equation}
V_{\delta}=\sum_{i\in
Z_{N}}e_{\delta(i)}e_{i}^{\dagger}\equiv\sum_{i\in
Z_{N}}(e_{i_{1}}\otimes e_{i_{1}})e_{i}^{\dagger},
\end{equation}
and its inverse
\begin{equation}
V_{\delta}^{\dagger}=V_{\delta^{-1}}=\sum_{i_{1}\in
Z_{N_{1}},i_{2}\in
Z_{N_{2}}}e_{\delta^{-1}(i_{1},i_{2})}e_{(i_{1},i_{2})}^{\dagger}\equiv
\sum_{i_{1}\in Z_{N_{1}},i_{2}\in Z_{N_{2}}}e_{\delta^{-1}(i_{1},i_{2}%
)}(e_{i_{1}}^{\dagger}\otimes e_{i_{1}}^{\dagger}),
\end{equation}
for which we have $V_{\delta}V_{\delta}^{\dagger}=V_{\delta}V_{\delta^{-1}%
}=V_{\delta\circ\delta^{-1}}=V_{id\times id}=\mathbf{1}_{\mathbf{C}_{N_{1}}%
}\otimes\mathbf{1}_{\mathbf{C}_{N_{2}}},$ and
$V_{\delta}^{\dagger}V_{\delta
}=V_{\delta^{-1}}V_{\delta}=V_{\delta^{-1}\circ\delta}=V_{id}=\mathbf{1}%
_{\mathbf{C}_{N}}.$

\bigskip

\textit{Notation:} in the sequel and in order to distinguish the
space dimensionality $N,$ referring to certain e.g functional,
operator, probability
vector, co-multiplication etc, we will denote it by $\phi_{\lbrack N]}%
,T_{[N]},$ and $P_{[N]}$ respectively. We can now state two
necessary and sufficient conditions, in
order to obtain an isomorphic \textit{prime decomposition }of a $\mathbf{Z}%
_{N}$ ARW governed by a pd $P_{[N]},$ into a product of two others
ARWs $\mathbf{Z}_{N_{1}},$ $\mathbf{Z}_{N_{2}},$ with respective
pd's $P_{[N]}=P_{[N_{1}]}\otimes P_{[N_{2}]}$. The first condition
is number theoretic and is about the compositeness of the
dimension number $N$ of the probability distribution $P_{[N]}$
that generates the ARW on $\mathbf{Z}_{N}$, while the second
condition is about its factorization into a tensor product of two
pd's of appropriate dimensions.

\textit{Proposition 3. i) }Let $N=N_{1}N_{2}$ be the prime
factorization of a positive integer $N,$ $\ $then if in addition
to the isomorphism of abelian groups of $Z_{N}\approx
Z_{N_{1}}\otimes Z_{N_{2}},$ we consider a probability
distribution (pd) $P_{[N]}$ factorizable into a product of two
others pd's
$P_{[N_{1}]},P_{[N_{2}]},$ namely such that $V_{\delta}P_{[N]}=P_{[N_{1}%
]}\otimes P_{[N_{2}]},$ or $p_{i}^{[N]}=p_{i_{1}}^{[N_{1}]}p_{i_{2}}^{[N_{2}%
]},$ then the functional of a $\mathbf{Z}_{N}$ algebraic random
walk $\phi_{\lbrack N]}=\sum_{i\in Z_{N}}p_{i}^{[N]}g_{[N]}^{i},$
with $0\leq p_{i}^{[N]}\leq1,\sum_{i\in Z_{N}}p_{i}^{[N]}=1,$
factorizes for every step $n,$ namely $\phi_{\lbrack N]}^{\ast
n}\approx\phi_{\lbrack N_{1}]}^{\ast n}\otimes\phi_{\lbrack
N_{2}]}^{\ast n},$ and similar factorization is valid for the
transition operator i.e $T_{[N]}^{n}\approx T_{[N_{1}]}^{n}\otimes
T_{[N_{2}]}^{n}.$

\textit{ii)} Let $N_{1},$ $N_{2}$ and $N_{3,}$ be the co-prime
factors of some positive integer $N$, then the decomposition
$Z_{N}\approx Z_{N_{1}}\otimes Z_{N_{2}}\otimes Z_{N_{3}}\approx
Z_{N_{1}N_{2}}\otimes Z_{N_{3}}\approx Z_{N_{1}}\otimes
Z_{N_{2}N_{3}},$ is co-assosiative\cite{abe} as indicated in the
last two equations, and if a pd $P_{[N]}$ is considered for which
the finest decomposition $V_{\delta}P_{[N]}=P_{[N_{1}]}\otimes
P_{[N_{2}]}\otimes P_{[N_{3}]},$ in terms of three others pd's is
true, then the associated $\mathbf{Z}_{N}$ ARW is also decomposed
at any step $n,$ namely for its functional and transition operator
respectively, the following co-associative factorizations are
valid
\begin{equation}
\phi_{\lbrack N]}\approx\phi_{\lbrack N_{1}]}\otimes \phi_{\lbrack
N_{2}]}\otimes\phi_{\lbrack N_{3}]}\approx \phi_{\lbrack
N_{1}N_{2}]}\otimes\phi_{\lbrack N_{3}]} \approx\phi_{\lbrack
N_{1}]}\otimes\phi_{\lbrack N_{2}N_{3}]},
\end{equation}

\textit{Proof: i) }The $N=N_{1}N$\bigskip$_{2}$ case: By means of
the factorization property of \ the pd viz.%
\begin{align}
V_{\delta}P_{[N]}  &  =\sum_{i\in
Z_{N}}p_{i}^{[N]}V_{\delta}e_{i}\approx
\sum_{i\in Z_{N}}p_{\delta(i)}^{[N]}e_{\delta(i)}=\sum_{i_{1}\in Z_{N_{1}%
},i_{2}\in Z_{N_{2}}}p_{(i_{1},i_{2})}^{[N]}e_{i_{1}}\otimes e_{i_{2}%
}\nonumber\\
&  =\sum_{i_{1}\in Z_{N_{1}},i_{2}\in Z_{N_{2}}}p_{i_{1}}^{[N_{1}]}p_{i_{2}%
}^{[N_{2}]}e_{i_{1}}\otimes e_{i_{2}}=P_{[N_{1}]}\otimes
P_{[N_{2}]},
\end{align}
we deduce that
\begin{align}
V_{\delta}D_{[N]}V_{\delta}^{\dagger}  &  =\sum_{i,j\in Z_{N}}p_{(i-j)}%
^{[N]}V_{\delta}e_{i}e_{j}^{\dagger}V_{\delta}^{\dagger}\nonumber\\
&  \approx\sum_{i_{1}\in Z_{N_{1}},i_{2}\in Z_{N_{2}}}p_{\delta(i-j)}%
^{[N]}e_{i_{1}}\otimes e_{i_{2}}.e_{i_{1}}^{\dagger}\otimes e_{i_{2}}%
^{\dagger}=D_{[N_{1}]}\otimes D_{[N_{2}]},
\end{align}
and therefore $V_{\delta}D_{[N]}^{n}V_{\delta}^{\dagger}\approx D_{[N_{1}%
]}^{n}\otimes D_{[N_{2}]}^{n},$ for the delta matrix. This yields
the factorization of the functional
\bea
 \phi_{\lbrack N]}&=&
G_{[N]}^{T}P_{[N]}=G_{[N]}^{T}V_{\delta
}^{\dagger}V_{\delta}P_{[N]}\nonumber\\
&\approx& (V_{\delta}G_{[N]}^{T})^{\dagger}(V_{\delta}P_{[N]})=G_{[N_{1}]}%
^{T}{}^{\dagger}\otimes
G_{[N_{2}]}^{T}{}^{\dagger}.P_{[N_{1}]}\otimes
P_{[N_{2}]}=\phi_{\lbrack N_{1}]}\otimes\phi_{\lbrack N_{2}]},
\eea

\nd and similarly for the transition operator (with $\tau(x\otimes
y)=y\otimes x$ ) \bea
 T_{[N]}  &=& (id\otimes\tau\otimes
id)\circ(\phi_{\lbrack N]}\otimes
id_{[N]})\circ\Delta_{\lbrack N]}\nonumber\\
&\approx& (\phi_{\lbrack N_{1}]}\otimes id_{[N_{1}]}\otimes
\phi_{\lbrack N_{2}]}\otimes id_{[N_{1}]})\circ(\Delta_{\lbrack N_{1}%
]}\otimes\Delta_{\lbrack N_{1}]})\approx T_{[N_{1}]}\otimes
T_{[N_{2}]}.\qquad
\eea

\bigskip\textit{ii)} The $N=N_{1}N_{2}N_{3}$ case: Applying the
CRT to $N,$ in two different ways yields for the abelian group
$Z_{N}$ two
factorizations i.e $Z_{N}\approx Z_{N_{1}}\otimes Z_{N_{2}}\otimes Z_{N_{3}%
}\approx Z_{N_{1}N_{2}}\otimes Z_{N_{3}}\approx Z_{N_{1}}\otimes
Z_{N_{2} N_{3}}.$ This can equivalently be expressed by means of
the bijection $\delta,$ as
\begin{equation}
(\delta_{\lbrack N_{1},N_{2}]}\otimes id)\circ\delta_{\lbrack N_{1}N_{2}%
,N_{3}]}=(id\otimes\delta_{\lbrack
N_{2},N_{3}]})\circ\delta_{\lbrack N_{1},N_{2}N_{3}]}.
\end{equation}
Performing the factorization of the pd $P_{[N]},$ as before twice
we have
\begin{equation}
(V_{\delta_{\lbrack
N_{1},N_{2}]}}\otimes\mathbf{1})V_{\delta_{\lbrack
N_{1}N_{2},N_{3}]}}P_{[N]}=P_{[N_{1}]}\otimes P_{[N_{2}]}\otimes P_{[N_{3}]}%
\end{equation}
which implies that%

\begin{equation}
(V_{\delta_{\lbrack
N_{1},N_{2}]}}\otimes\mathbf{1})V_{\delta_{\lbrack
N_{1}N_{2},N_{3}]}}D_{[N]}^{n}((V_{\delta_{\lbrack N_{1},N_{2}]}}%
\otimes\mathbf{1})V_{\delta_{\lbrack
N_{1}N_{2},N_{3}]}})^{\dagger}\approx D_{[N_{1}]}^{n}\otimes
D_{[N_{2}]}^{n}\otimes D_{[N_{3}]}^{n}.
\end{equation}
This factorization of the delta matrix leads to an equivalent
factorizations of the functional at each time step of the
$\mathbf{Z}_{N}$ walk, i.e
\begin{equation}
\phi_{\lbrack N]}^{\ast n}\approx\phi_{\lbrack N_{1}N_{2}]}^{\ast n}%
\otimes\phi_{\lbrack N_{3}]}^{\ast n}\approx\phi_{\lbrack
N_{1}]}^{\ast n}\otimes\phi_{\lbrack N_{2}N_{3}]}^{\ast
n}\approx\phi_{\lbrack N_{1}]}^{\ast
n}\otimes\phi_{\lbrack N_{2}]}^{\ast n}\otimes\phi_{\lbrack N_{3}]}^{\ast n}%
\end{equation}
Similar decompositions can be obtained for Markov transition
operators. \ \ \ \ \ \ \ \ \ \ \ \ \ \ \ \ \ \ \ \ \ \ \ \ \ \ \ \
\ \ \ \ \ \ \ \ \ \ \ \ \ \ \ \ \ \ \ \ \ \ \ \ \ \ \ \ \ \
$\square $

\textit{Remark:} Interpretation of this result implies that the
original random walk $\mathbf{Z}_{N},$ is isomorphically
decomposed into the product of : i) three similar walks
$\mathbf{Z}_{N}\cong\mathbf{Z}_{N_{1}}\otimes
\mathbf{Z}_{N_{2}}\otimes\mathbf{Z}_{N_{3}},$ ii) or into the
product of two
walks of dimensions $N_{1}N_{2}$ and $N_{3}$ i.e $\mathbf{Z}_{N_{1}N_{2}%
}\otimes\mathbf{Z}_{N_{3}}$ and iii) or into the product of two
walks of dimensions $N_{1}$ and $N_{2}N_{3}$ i.e
$\mathbf{Z}_{N_{1}}\otimes \mathbf{Z}_{N_{2}N_{3}}.$ As an example
we take the \ $\mathbf{Z}_{6}$ random walk on the vertices of \ a
canonical hexagon\ the statistics of which is determined by a
$6$-dimension pd vector $P_{[6]}.$ If this pd has been chosen so
that there are two other pd's one 3-dimensional $P_{[3]}$ that
generates a $\mathbf{Z}_{3}$ ARW on the vertices of a canonical
triangle, and one 2-dimensional \ $P_{[2]}$ that generates a
$\mathbf{Z}_{2}$ walk on two points, such that \
$P_{[6]}=P_{[3]}\bigotimes P_{[2]},$ then we can decompose \ the
ARW on the canonical hexagon as a product of two walks one \ on
the triangle times one on the two-point set.\ Similar
interpretations can be given to the prime decomposition of an ARW
on\ \ a canonical polygon. E. g \ the
decomposition $\mathbf{Z}_{60}\cong$ $\mathbf{Z}_{3}\bigotimes\mathbf{Z}%
_{4}\bigotimes\mathbf{Z}_{5}\cong$ $\mathbf{Z}_{12}\bigotimes\mathbf{Z}%
_{5}\cong$
$\mathbf{Z}_{3}\bigotimes\mathbf{Z}_{20}$\cite{ellinasf}.\ \ \ \ \
\ \ \ \ \ \ \ \ \ \ \ \ \ \ \ \ \ \ \ \ \ \ \ \ \ \ \ \ \ \ \ \ \
\ \ \ \ \ \ \ \ \ \ \ \ \ \ \ \ \ \ \ \ \ \ \ \ \ \ \ \ \ \ \ \ \
\ \ \ \ \ \ \ \ \ \ \ \ \ \ \ \ \ \ \ \ \ \ \ \ \ \ \ \ \ \ \ \ \
\ \ \ \ \ \ \ \ \ \ \ \ \ \ \ \ \ \ \ \ \ \ \ \ \ \ \ \ \ \ \

\textit{Problem: }Let $N=N_{1}N_{2}$ be the prime factorization of
a positive integer $N,$ for which the isomorphism $Z_{N}\approx
Z_{N_{1}}\otimes Z_{N_{2}},$ is valid, let us consider an
$\mathbf{Z}_{N}$ ARW generated as previously by a pd $P_{[N]}.$ It
this pd is not factorizable into a product of two others pd's as
in the remark above, but instead there are two pairs of pd's
$P_{[N_{1}]},$ $P_{[N_{1}]}^{/},$ and $P_{[N_{2}]},$
$P_{[N_{2}]}^{/},$ such that the original pd is a convex
combination of them i. e $(0\leq q\leq1),$
\begin{equation}
V_{\delta}P_{[N]}=qP_{[N_{1}]}\otimes
P_{[N_{2}]}+(1-q)P_{[N_{1}]}^{/}\otimes P_{[N_{2}]}^{/}.
\end{equation}

In the terminology of quantum information\cite{nielsen}, we have
here two ARWs $\mathbf{Z}_{N_{1}},\mathbf{Z}_{N_{2}},$ that are
\textit{classically correlated (cc) }and form a probabilistic
decomposition of the ARW in $\mathbf{Z}_{N},$ and write
symbolically $\mathbf{Z}_{N}\cong q\mathbf{Z}
_{N_{1}}\bigotimes\mathbf{Z}_{N_{2}}+(1-q)\mathbf{Z}_{N_{1}}^{/}%
\bigotimes\mathbf{Z}_{N_{2}}^{/}.$ A number of interesting
problems arise in this context: the total dynamics and reduced
dynamics of the components of the $\mathbf{Z}_{N}$ walk; the
problem of construction of measures of correlations among the
components of $\mathbf{Z}_{N}$ walk; the problem of
information(majorization) dynamics and information exchange among
the pd's components of the $\mathbf{Z}_{N}$ walk; the problem of
asymptotics of the $\mathbf{Z}_{N}$ walk.

\subsection{The case of {\bf hw}-Algebra}
A bialgebra \cite{abe} ${\cal A}={\cal A}(\mu , \eta, \D , \e)$
over a field $k$ is a vector space equipped with an algebra
structure with homomorphic associative product map
$\mu:\aa\times\aa\rightarrow\aa$, and a homomorphic unit map
$\eta:k \rightarrow \aa$, that are related by
$\mu\circ(\eta\otimes id)=id=\mu\circ(id\otimes\eta)$, together
with a coalgebra structure with a homomorphic coassociative
coproduct map $\D:\aa\rightarrow\aa\otimes\aa$ and a homomorphic
counit map $\e:\aa\rightarrow k$, that are related between them by
$(\e\otimes id)\circ\D = id = (id\otimes \e)\circ\D $. Both
products satisfy the compatibility condition of bialgebra i.e
$(\mu\otimes\mu)\circ(id\otimes\tau\otimes id)\circ(\D\otimes\D)=
\D\circ\mu $, where $\tau(x\otimes y)=y\otimes x$ stands for the
twist map.
If $\eta$ or $\e$ is not defined in $\aa$ we speak about non
unital or non counital Hopf algebra.

Suppose we have a functional $\phi:\aa\rightarrow \bf C$, defined
on $\aa$, let us define the operator $T_\phi :\aa \rightarrow \aa$
as $T_\phi =(\phi \otimes id )\circ \D$, then $\e\circ T_\phi =
\phi$, namely the counit aids to pass from the operator to its
associated functional. From this relation we can define the
convolution product $\psi * \phi $, between functionals as follows
\cite{majid1}: \bea \e\circ T_\psi T_\phi &=& \e \circ (\psi
\otimes id )\circ \D \circ (\phi \otimes id )\circ \D \nn \\ &=&
(\phi\otimes\psi )\circ
(id\otimes id\otimes\e )\circ (id \otimes \D )\circ \D \nn \\
&=& (\phi\otimes\psi )\circ \D = \phi*\psi \;, \eea

\nd  and in general $\e \circ T_{\phi}^n =\e \circ
T_{\phi^{*n}}=\phi^{*n}$. These last relations imply that the
transition operators form a discrete semigroup with respect to
their composition with identity element $T_\e \equiv id$ (due to
the axioms of bialgebra) and generator $T_\phi$, while the
functionals form a dual semigroup with respect to the convolution
with identity element $e$ and generator $\phi$, and that these two
semigroups are homomorphic to each other.

We recall now the {\it Heisenberg-Weyl algebra} $\bf hw$ and its
structural maps: this is the algebra of the quantum mechanical
oscillator and is generated by the creation, annihilation and the
unit operator $\{ \ad , a, \en \}$ respectively which satisfy the
commutation relation (Lie bracket) $[a, \ad ]=\en $, while $\en$
commutes with the other elements. This algebra possesses a non
counital bialgebra structure \cite{majidbook}, chapt. 3), with
comultiplication defined as \bea \D^{(n-1)}a &=&
n^{-\frac{1}{2}}(a\otimes\cdots \otimes \en + \en\otimes a\otimes
\cdots \otimes \en +
\en \otimes \cdots \otimes a)\;, \nn \\
\D^{(n-1)}\ad &=& n^{-\frac{1}{2}}(\ad\otimes\cdots \otimes \en +
\en \otimes a\otimes \cdots \otimes \en +
\en \otimes \cdots \otimes \ad)\;, \nn \\
\D \en &=& \en \otimes \en \;, \label{comult} \eea

\nd where as indicated above the $\D^{(n-1)}$ maps the
creation/annihilation operators into the $n$th fold tensor product
of the algebra and adds appropriate factors (also c.f.
\cite{hudson}). Let us also define the number operator $N=\ad a$
with the following commutation relations with the generators of
$\bf hw$:
$[N,\ad ]=\ad  \;, [N,a ]= -a $.
The module which carries the unique irreducible and infinite
dimensional representation of the oscillator algebra is the
Hilbert-Fock space ${\cal H}_F$ which is generated by a lowest (or
"vacuum" ) state vector $\ket{0}\in \cal H_F$ and is given as
${\cal H}_F =\{\ket{n}=\frac{(\ad)^{n} }{n!}\ket{0}, n\in {\bf
Z}_+ \}$.

The functionals we intend to use will be defined by means of the
canonical coherent state vectors of the $\bf hw$ algebra so in the
sequel we give a brief introduction to the concept of coherent
state vectors (CSV) on Lie groups: consider a Lie group $\cal G$,
with a unitary irreducible representation $T(g)$, $g\in \cal G$,
in a Hilbert space $\cal H$. We select a reference vector
$\ket{\Psi_0}\in \cal H$, to be called the "vacuum" state vector,
and let ${\cal G}_0\subset \cal G$ be its isotropy subgroup, i.e
for $h\in {\cal G}_0$, $T(h)\ket{\Psi_0}=
e^{i\varphi(h)}\ket{\Psi_0}$. The map from the factor group ${\cal
M} = {\cal G}/{\cal G}_{0}$  to the Hilbert space $\cal H$,
introduced in the form of an orbit of the vacuum state under a
factor group element, defines a CSV $\ket{x}=T({\cal G}/{\cal
G}_{0})\ket{\Psi_0}$ labelled by points $x\in\cal M$ of the
coherent state manifold. Coherent states form an (over)complete
set of states, since by means of the Haar invariant measure of the
group $\cal G$ \viz $\mes{x}, \;\; x\in \cal M$, they provide a
resolution of unity, $\bf{1}=\it{\int_{\cal
M}\mes{x}\proj{x}{x}}$. As a consequence, any vector
$\ket{\Psi}\in\cal H$ is analyzed in the CS basis,
$\ket{\Psi}=\int_{\cal M}\mes{x}\Psi(x)\ket{x}$, with coefficients
$\Psi(x)=\inner{x}{\Psi}$. We should note here that the square
integrability of the vectors ${\Psi}$ will impose some limits on
the growth parameters of the functions $\Psi(x)$ at the boundary
of manifold $\cal M$ (cf. \cite{klauder,perelomov} and references
therein).


The $\bf hw$-CS is defined by the relation \be
\ket{\a}=e^{\a\ad-\ab a}\ket{0}= {\cal N}
e^{\a\ad}\ket{0}=e^{-{\fscr{1}{2}}\absqr{\a}}\sum_{n=0}^{\infty}
\frac{\a^n}{\sqrt{n!}}\ket{n}\;. \ee

\nd It is an (over)complete set of normalized states with respect
to the measure $\mes{\a}=\frac{1}{\pi}e^{-\absqr{\a}} d^{2}\a$ for
the non-normalized CS, and $\a\in{\cal M}= HW/U(1)\approx \bf C$
is the CS manifold. Since $a\ket{\a}=\a\ket{\a}$, $\cal M$ is the
flat canonical phase plane with the standard line element $ds^2
=d\a d\ab$. Also the symplectic 2-form $\om=id\a \wedge d\ab$ is
associated to the canonical Poisson bracket $\{f,g\}=i(\pl{\a} f
\pl{\ab} g - \pl{\ab} f \pl{\a} g) $.

The density operator (state) $\rho$ which would be used to
determine functionals of some operator bialgebras $\cal A$, is
defined generally  as follows : Let a Hilbert vector space $\cal
H$ that carries a unitary irreducible representation of $\cal A$
of finite or infinite dimension. The set of density operators \be
{\cal S}=\{\rho \in {\rm End}({\cal H}): \rho\geq 0, \rho^\dag
=\rho, tr\rho=1\}\;, \ee

\nd namely the set of non-negative, Hermitian, trace-one operators
acting on $\cal H$ form a convex subspace of  $ {\rm End}({\cal
H})$, which is the convex hull of the set \be {\cal S}_P = \{\rho
\in {\cal S}, \rho^2 =\rho \}\equiv {\cal H}/U(1)\;, \ee

\nd namely of the set of pure density operators (states), that are
in one-to-one correspondence with the state vectors of $\cal H$.

\nd The density operator to be used in the case of $\bf hw$ walk
uses the pure density operators $\proj{\pm \a}{\pm \a}\in {\cal
S}_P $ and is a convex combination belonging to the convex hull of
${\cal S}_P$ i.e $0\leq p\leq 1$, \bea \rho &=& p\proj{\a}{\a}
+(1-p)\proj{-\a}{-\a} \;.\eea

\nd  Let $\phi(\cdot)={\rm Tr}\rho(\cdot)\equiv<\rho ,\cdot
>$, a functional defined on the enveloping Heisenberg-Weyl algebra
${\cal U}(hw)$, where $\rho=p\proj{\a}{\a}+(1-p)\proj{-\a}{-\a}$,
i.e the $\rho$ density operator is given as a convex sum of pure
state density operators. The action of the transition operator
$T_\phi = (\phi \otimes id)\circ \D $ on the generating monomials
of ${\cal U}(hw)$ (where we ignore the numerical factors in the
comultiplication of eq.(\ref{comult})) reads, \bea
T_\phi ((\ad)^{m} a^{n} ) &=& (\phi \otimes id)\circ \D ( (\ad)^{m} a^{n} )\nn \\
&=& \sum_{i=0}^{m} \sum_{j=0}^n \col{m}{i}\col{n}{j}
[p \a^{* i}\a^{j} + (1-p)(-\a)^{i} (-\a)^j ] (\ad)^{m-i}a^{n-j}\nn \\
&=& p (\ad + \a^{*} )^{m} (a + \a )^{n} + (1-p) (\ad - \a^{*} )^m
(a- \a )^n \;. \eea

\nd For a general element $f(a,\ad ) \in {\cal U}(hw)$ that is
normally ordered, namely the annihilation operator $a$ is placed
to the right of the creation operator $\ad$, denoted by $f (a,\ad
) =\sum_{m,n \geq 0}c_{mn} (\ad)^m a^n$, the action of the linear
operator $T_\phi$ becomes \be T_\phi  (f(a,\ad ) ) = p f(a + \a
,\ad +\a^* ) + (1-p) f(a - \a ,\ad -\a^* ) \ee


\nd By means of the CS eigenvector property and the normal
ordering of the $f$ element we also compute the value of
functional viz. \be \phi (f(a,\ad ))= p f( \a ,\a^* ) + (1-p) f( -
\a , -\a^* ) \;. \ee

Let us consider the displacement operator $D_\a = e^{\a \ad - \a^*
a}\equiv e^\Lambda$, with $\Lambda=\alpha
a^{\dagger}-\overline{\alpha}a$ which acts with the group adjoint
action on any element $f$ of the ${\cal U}(hw)$ algebra
viz.\cite{klauder} \be Ad D_a (f)=Ad e^{\a \ad - \a^* a}(f)=  e^{
ad(\a \ad - \a^* a)}(f)= e^{ad\Lambda}= D_\a f D_\a ^\dagger , \ee

\nd where $ad (X)f=[X,f]$ and $ad(X)ad(X)f =[X,[X,f]]$ and
similarly for higher powers, stands for the Lie algebra adjoint
action that is defined in terms of the Lie commutator. Similarly
the group adjoint action in terms of the displacement operator on
the generators of ${\cal U}(hw)$ reads $Ad D_{\pm \a}(a)=a\mp \a$
and $Ad D_{\pm \a}(\ad )=\ad \mp \a^* $. By means of these
expressions we rewrite the action of the transition operator as
\bea
 T_\phi (f(a,\ad )) &=&[p Ad D_{-\a} + (1-p) Ad D_{\a}]f(a,\ad)  \nn \\ &=& [ pe^{-ad\Lambda}+(1-p)e^{ad\Lambda}]f(a,\ad
 ).
\eea

Next we compute the limiting transition operator
 \begin{align}
T_{t}&\equiv
T_{\phi_{t}}\equiv\lim_{n\rightarrow\infty}T_{\phi}^{n}= \nn \\
&\lim_{n\rightarrow\infty}\Bigl[ 1+\frac{t}{n}ad\Lambda +
\frac{t}{n} \gamma(ada^{\dagger})^{2} +
\frac{t}{n}\overline{\gamma}(ada)^{2}- \nn \\
&\hspace{5cm} \frac{t} {n}\left\vert \gamma\right\vert (ada\text{
}ada^{\dagger}+ada^{\dagger}\text{ }ada^{\dagger})\Bigl]  ^{n}.
\end{align}

\nd In the last expression we have introduced the parameters of
continuous time $t\in\mathbf{R}$ and the drift and diffusion terms
respectively
$c,\gamma\in\mathbf{C}$ by means of the relations%
\begin{equation}
2\alpha(p-\frac{1}{2})=\frac{tc}{n},\text{
}\frac{a^{2}}{2}=\frac{t\gamma}{n},
\end{equation}
and have performed the limits $\alpha\rightarrow0,$
$n\rightarrow\infty,$ with $t,$ $c,$ $\gamma$ been fixed. We have
also been used the limit
lim$_{n\rightarrow\infty}(1+\frac{Z}{n})=e^{Z},$ to obtain the
continuous time  Markov transition operator $T_{t}=e^{tL},$ with
its $L$ generators as it is obviously identified in the equation
below

\begin{equation}
T_{t}=e^{tL}\equiv\exp t\left[  1+ad\Lambda+\gamma(ada^{\dagger}%
)^{2}+\overline{\gamma}(ada)^{2}-\left\vert \gamma\right\vert
(ada\text{ }ada^{\dagger}+ada^{\dagger}\text{
}ada^{\dagger})\right]  .
\end{equation}

By construction $T_t$ is the time evolution operator for any
element  $f$ of ${\cal U}(hw)$ i.e $f_t =T_t (f)$ and forms a
continuous semigroup $T_t T_{t'} = T_{t+t'}$ under composition.
This yields the diffusion equation obeyed by  $f_t$, which will be
taken to be normally ordered hereafter. By time derivation of the
equation \be \phi_t (f)=<\rho , f_t>= <\rho , e^{t ad {\cal
L}}f>=< e^{-t ad{\cal L}^\dagger }\rho , f> =<\rho_t , f>\;, \ee

\nd we obtain the diffusion equation $\frac{d}{dt}f_t  ={\cal
L}f_t  $, as well as the dual one satisfied by the $\rho$ density
operator viz. $\frac{d}{dt}\rho_t = {\cal L}^\dagger \rho_t $.

Explicitly the quantum master evolution equation for the density
matrix reads
\begin{align}
\frac{d}{dt}\rho(t)  &
=[ca^{\dagger}-\overline{c}a,\rho]+\gamma(a^{\dagger
2}\rho+\rho a^{\dagger}-2a^{\dagger}\rho a^{\dagger})+\overline{\gamma}%
(a^{2}\rho+\rho a^{2}-2a\rho a)\nonumber\\
&  -\left\vert \gamma\right\vert \left(  (2N+1\right)  \rho+\rho
(2N+1)-2a^{\dagger}\rho a-2a\rho a^{\dagger}).
\end{align}

The obtained equation is similar to the quantum master equation
that describes the trace preserving dynamics of the reduced
density matrix operator of single mode of the electromagnetic
field interacting coherently with classical electric filed while
it is immersed in a bath of quantum oscillators\cite{scully}. The
decay of the field mode is influenced by the kind of initial
condition the reservoir oscillators are put in. To analyze the
physical content of that equation we rewrite it below by
separating its right hand side into three lines i.e
\begin{align}
\frac{d}{dt}\rho(t)  &  =[ca^{\dagger}-\overline{c}a,\rho]\nonumber\\
&  -\left\vert \gamma\right\vert (a^{\dagger}a\rho+\rho
a^{\dagger}a-2a\rho
a^{\dagger}+aa^{\dagger}\rho+\rho aa^{\dagger}-2a^{\dagger}\rho a)\nonumber\\
&  +\gamma(a^{\dagger2}\rho+\rho a^{\dagger}-2a^{\dagger}\rho
a^{\dagger })+\overline{\gamma}(a^{2}\rho+\rho a^{2}-2a\rho a).
\end{align}
The first line gives the coherent interaction of the mode with the
classical electric filed of intensity $c$, as described by the
commutator of density operator with the Hamiltonian term. It is
neglected for balanced walk. The second line is a typical part of
a master equation describing mode decaying for reservoir
oscillators in thermal equilibrium\cite{comment}. The last line is
related to the case where the reservoir is prepared in a squeezed
vacuum state\cite{scully}. Closing we should notice that the above
quantum master equation can be transformed into a Fokker-Planck
partial differential equation for some quasi-probability function
e.g $P, Q$, or Wigner function associated with the density
operator e.g \cite{scully}.

\section{Quantum Random Walks}
\subsection{The case of {\bf Z}}
In a quantum random walk there are two dynamically coupled
systems: the walker systems described by a Hilbert space $H_{w},$
and the coin system
also described by a 2D Hilbert space $H_{c}\approx\mathbf{C}^{2}%
=$span$(|+>,|->).$ Let the unitary matrix $U$ operating in
$H_{c},$ e.g $U_{H}=\frac{1}{\sqrt{2}}\left(
\begin{array}
[c]{cc}%
1 & 1\\
1 & -1
\end{array}
\right)$, or $U_{\frac{\pi}{4}}=\sigma_{3}U_{H},$ the Hadamard
(Fourier) transform and \ the $\frac{\pi}{4}-$ rotation matrix
respectively. If we denote by $P_{\pm}=|\pm><\mp|,$ the projection
of an orthogonal partition of $H_{c},$ and by $S_{\mp},$ two step
operators in the walker's space $H_{w}$ (explicit examples
determined below), we introduce the unitary one-step evolution
operator acting in the space $H_{c}\otimes H_{w}$ in the combined
coin+walker system:%

\begin{equation}
V=\sum_{m=\pm}P_{m}U\otimes S_{m}=\frac{1}{\sqrt{2}}\left(
\begin{array}
[c]{cc}%
S_{+} & S_{+}\\
\pm S_{-} & \mp S_{-}%
\end{array}
\right)  .
\end{equation}

In the equation above the upper/lower signs correspond to the
choices $U_{H},U_{_{\frac{\pi}{4}}},$ respectively. If initially
the two systems are
decoupled, their density matrices are factorized i.e $\varrho_{c}%
\otimes\varrho_{w}.$ Then if we assume that
$\rho_{c}=|\Psi><\Psi|,$ is a projective density matrix with
$|\Psi>=a|+>+b|->$, a normalized coin state vector, then the
1-time step of the QRW is considered as a completely positive
trace (CPTP) map $\varepsilon_{v},$ operating on the walkers'
density
operator, obtained by partially tracing out ("forgetting") the coin system i.e%
\begin{equation}
\varepsilon_{V}(\varrho_{w})=Tr_{c}(V\varrho_{c}\otimes\varrho_{w}V^{\dagger
})=\sum_{m=\pm}p_{m}S_{m}\varrho_{w}S_{m}^{\dagger}.
\end{equation}
Unitarity of $V,\ $\ implies that the two probabilities
$p_{\pm}=TrP_{\pm
}U\rho_{c}U^{\dagger}P_{\pm}=\frac{1}{2}\pm2\operatorname{Re}a\overline{b}%
\geq0,$ $p_{+}+p_{-}=1,$ are determined by the coin system state
vector variables and in turn they determine the Kraus
generator\cite{kraus}, ($\sqrt{p_{+}}S_{+},\sqrt{p_{-}}S_{-}),$ of
CPTP evolution map. Four sources of choices are implicit in the
above prescription of QRW: the choice of the initial coin+walker
state vectors, the choice of unitary $U,$ the choice of definition
of time step in terms of the tracing of the coin system (to be
investigated in detail below), and finally the choice of step
operators $S_{\mp}$ in the walker's space, that determines the
king of the QRW under investigation.

Next we choose to turn to a Hadamard random walk on integers with
dynamical algebra the Euclidean algebra
$\mathbf{e}(2)\thickapprox\mathbf{iso}(2)$, with
step operators $(\sqrt{p_{+}}S_{+},\sqrt{p_{-}}S_{-})=(\sqrt{p}E_{+}%
,\sqrt{(1-p)}E_{-}),$ and third element the "distance" operator
$L.$ The latter is the interesting quantum observable the quantum
moments of which are used to compare classical and quantum walks.
We examine three possible tracing schemes: i) the classical scheme
that \textit{promptly traces the coin system}
after each $V$ action, leading to the CRW%
\begin{equation}
\varepsilon_{V}(\rho_{w}^{(n)})=Tr_{c}(V\rho_{c}\otimes\rho_{w}^{(n-1)}%
V^{\dagger})=pE_{+}\varrho_{w}^{(n-1)}E_{+}^{\dagger}+(1-p)E_{-}\varrho
_{w}^{(n-1)}E_{-}^{\dagger},
\end{equation}
which produces the diagonal sequence of density matrices%
\begin{equation}
\left\{  \rho_{w}^{(0)},\varepsilon_{V}(\rho_{w}^{(0)}),\varepsilon_{V}%
^{2}(\rho_{w}^{(0)}),\varepsilon_{V}^{3}(\rho_{w}^{(0)}),\varepsilon_{V}%
^{4}(\rho_{w}^{(0)}),...\right\}  ,
\end{equation}
with diagonal elements the probabilities of site occupancy, given
by the $(p,1-p)$ classical Pascal triangle, ii) the scheme that
\textit{traces the coin system by increasing delays} i.e \ after
an increasing number of actions of $V$ operator, leading to a walk
designated by QRW$_{1}$
\begin{equation}
\bigskip\varepsilon_{V^{N}}(\rho_{w}^{(0)})=Tr_{c}(V^{N}\rho_{c}\otimes
\rho_{w}^{(0)}V^{\dagger N})=\sum_{m=\pm}S_{m}^{(N)}\varrho_{w}^{(0)}%
S_{m}^{(N)\dagger},
\end{equation}
which produces the sequence of density matrices%

\begin{equation}
\left\{  \rho_{w}^{(0)},\varepsilon_{V}(\rho_{w}^{(0)}),\varepsilon_{V^{2}%
}(\rho_{w}^{(0)}),\varepsilon_{V^{3}}(\rho_{w}^{(0)}),\varepsilon_{V^{4}}%
(\rho_{w}^{(0)}),...\right\}  ,
\end{equation}
iii) and the scheme of \textit{delaying the trace of coin system
by exactly one action of }$V$\textit{ operator, }leading to a walk
designated by
QRW$_{2}$%

\begin{equation}
\varepsilon_{V^{2}}(\rho_{w}^{(n)})=Tr_{c}(V^{2}\rho_{c}\otimes\rho
_{w}^{(n-1)}V^{2\dagger})=B_{+}\varrho_{w}^{(n-1)}B_{+}^{\dagger}+B_{-}%
\varrho_{w}^{(n-1)}B_{-}^{\dagger},
\end{equation}
which produces the sequence of density matrices%
\begin{equation}
\left\{
\rho_{w}^{(0)},\varepsilon_{V^{2}}(\rho_{w}^{(0)}),\varepsilon^{2}
_{V^{2}}(\rho_{w}^{(0)}),\varepsilon^{3}_{V^{2}}(\rho_{w}^{(0)}),\varepsilon^{4}
_{V^{2}}(\rho_{w}^{(0)}),...\right\}  \text{ .}%
\end{equation}

Next proposition deals with the time evolution of the pd \bigskip$(P_{q}%
^{(N)})_{m}=\left\langle m\right\vert \rho_{w}^{(N)}\left\vert
m\right\rangle ,$ made of the diagonal elements of the density
matrices in the course of the walk QRW$_{1}$ and
QRW$_{2}$\cite{bet}.

\textit{Remarks:} 1) From the above treatment can been shown that
the QRW$_1$ reproduces along the diagonal elements of the sequence
of evolving density matrices of the walker system the pd of the 1D
Hadamard random walk and its diffusion rate\cite{bet} 2) by
expressing the evolution unitary operator $V$, as the exponential
of a hermitian operator $H$ i.e $V=e^{iH}$, which describes an
interaction between coin and walker quantum systems, the various
schemes of partial tracing can be physically implemented by
choosing the length of interaction time e.g the QRW$_2$ scheme
requires an interaction time $t=2$ for the evolution operator $V_t
=\exp{(itH)}$.\cite{ellinasetal}

\textit{Proposition }\textit{4. }There exists bistochastic
matrices
$\Delta_{q}=S_{0}\circ\overline{S}_{0}+S_{1}\circ\overline{S}_{1},$
and
$\Delta_{q}=B_{0}\circ\overline{B}_{0}+B_{1}\circ\overline{B}_{1},$which
determine the pd drawn from the diagonal elements of the evolving
density matrices of the walks QRW$_{1},$ and QRW$_{2}$
respectively, by means of \ the respective equations
\begin{align}
P_{q}^{(n+1)}  &  =\Delta_{c}P_{q}^{(n)}+(E_{+}-E_{-})M^{(n)}P_{q}^{(0)},\\
P_{q}^{(n+1)}  &  =\Delta_{q}P_{q}^{(n)},
\end{align}
where
\begin{equation}
M^{(n)}=\sqrt{p(1-p)}S_{0}^{(N)}\circ\overline{S}_{1}^{(N)}+S_{1}^{(n)}%
\circ\overline{S}_{0}^{(n)}+(2p-1)S_{0}^{(n)}\circ\overline{S}_{0}^{(n)},
\end{equation}
and%
\begin{align}
\Delta_{q}  &  =\left\vert pa+\sqrt{p(1-p)}b\right\vert ^{2}E_{+}%
^{2}+\left\vert \sqrt{p(1-p)}a-pb\right\vert ^{2}E_{-}^{2}\nonumber\\
&  +\left\vert (1-p)a-\sqrt{p(1-p)}b\right\vert ^{2}+\left\vert
(1-p)b+\sqrt
{p(1-p)}a\right\vert \mathbf{1.}%
\end{align}

\nd Above the element by element or Hadamard product
$\Delta=A\circ B,$ defined between matrices $A,$ $B$ of the same
size by $(A\circ B)_{ij}=A_{ij}B_{ij},$ has been
used\cite{bhatia}. The study of the pd obtained from the previous
proposition reveals two novel aspects of the models QRW$_{1}$ and
QRW$_{2}$. First the aspect of breaking the condition
\textit{majorization-implies- entropy increase}, and that of
\textit{enhanced diffusion rates}. Before closing this section we
give a brief demonstration of the latter one (detail investigation
together with relevant references can be found in \cite{bet}).

\textit{Enhanced Diffusion Rates:}Let us consider the $m$th order
statistical moment of the distance operator
$<L^{m}>_{n}=Tr(\rho_{w}^{(n)}L^{m})$ , at the $n$th step. Assume
we have a symmetric walk with \ $<L>_{n}=0,$ so that the standard
deviation at n$th$ step is $\sigma_{n}=\sqrt{<L^{2}>_{n}}.$ For
the CRW we have that $\sigma_{n}^{C}=\sqrt{n},$ and from the pd's
taken by the previous proposition we obtain the respective
standard deviation for the walks QRW$_{1}$ and QRW$_{2},$
expressed in terms of their classical counterparts, for the first
five steps:$\{\sigma_{1}^{QRW_{1}}=$ $\sigma_{1}^{C},\sigma
_{2}^{QRW_{1}}=\sigma_{2}^{C},\sigma_{3}^{QRW_{1}}=\sigma_{3}^{C},\sigma
_{4}^{QRW_{1}}=\sqrt{5}/2\sigma_{4}^{C},\sigma_{5}^{QRW_{1}}=\sqrt{8/5}%
\sigma_{5}^{C}$ $\},$ and
$\{\sigma_{1}^{QRW_{2}}=\sigma_{1}^{C},\sigma
_{2}^{QRW_{2}}=\sqrt{5/2}\sigma_{2}^{C},\sigma_{3}^{QRW_{3}}=\sqrt{3}%
\sigma_{3}^{C},\sigma_{2}^{QRW_{2}}=\sqrt{7/2}\sigma_{4}^{C},\sigma
_{5}^{QRW_{2}}=2\sigma_{5}^{C}\}.$

From that we deduce that there is a quadratic speed up of the
spreading rate in the case of QRW$_{1}$ with respect to CRW, and
that rate is even bigger for the case of QRW$_{2}.$ Finally the
asymptotic ($n\gg1),$ growth values are turn out to be
$\sigma_{n}^{QRW_{1}}\sim\sqrt{n(2-\sqrt{2})/2}$ $\sigma
_{n}^{C}.$

\section{Relation of Algebraic and Quantum Random Walks}
This section will put forward a relation between the two types of
random walks under investigation so far i.e the ARQ and QRW. The
relevant theory here is Naimark's extension theorem that allows to
express in a non unique manner a positive trace preserving map,
operating by means of its Kraus generators on a density matrix
describing the state of some quantum system in a certain Hilbert
space, by a unitary operator acting on a extension of the original
space.

Stated in the language of random walks the extension theorem
assumes a ARW
described by a CPTP map $\varepsilon(\varrho_{w})=\sum_{m=\pm}p_{m}%
S_{m}\varrho_{w}S_{m}^{\dagger},$ operating on the density matrix
of \textit{walker system }\ with its Kraus generators
$(\sqrt{p_{+}}S_{+} ,\sqrt{p_{-}}S_{-}),$ defined to act on
Hilbert space $H_{w}.$ It is further assumed as usually for ARWs,
that the step generators $S_{\pm}$ are related to and algebra of
operators that needs to be specified. Then a unitary operator $V$
is considered acting on $H_{c}\otimes H_{w}$ i.e an extension of
the
original space by an extra or ancilla space $H_{c}\approx\mathbf{C}^{2}%
=$span$(|+>,|->),$ which in the context of QRW stands for the
\textit{coin system}. Let a pure density matrix in the coin system
$\rho_{c}=|\Psi><\Psi|.$ Then the extension theorem provides a
unitary representation of the CPTP i.e
\begin{equation}
\varepsilon_{V}(\varrho_{w})=\sum_{m=\pm}p_{m}S_{m}\varrho_{w}S_{m}^{\dagger
}=Tr_{c}(V\varrho_{c}\otimes\varrho_{w}V^{\dagger}).
\end{equation}
The unitary operator provides the Kraus generators as $\sqrt{p_{m}}%
S_{m}=\left\langle m\right\vert V\left\vert \Psi\right\rangle ,$
up to a local unitary operator $W,$ i.e the transformation \
$V\rightarrow W\otimes \mathbf{1}_{w}\mathbf{\ }V,$ provides the
same generators. For the case of ARQs the unitary operator $V$ is
specifically expressed by means of the coin states projections
$P_{\pm},$ and the unitary matrix $U(p_{\pm})=\left(
\begin{array}
[c]{cc}%
\sqrt{p_{+}} & \sqrt{p_{-}}\\
\sqrt{p_{-}} & -\sqrt{p_{+}}%
\end{array}
\right)  $ of the coin space, as $V=\sum_{m=\pm}P_{m}U\otimes
S_{m}.$ In particular the step operators are unitary and inverse
to each other i.e $S_{+}=(S_{-})^{\dagger}.$ This unitary
representation can be extended to products of CPTM maps by first
defining the unitaries
\begin{equation}
V^{\otimes2}\equiv\sum_{m,n=\pm}P_{m}U\otimes P_{n}U\otimes
S_{m}S_{n}.
\end{equation}
\bigskip Then \ we obtain
\begin{equation}
\varepsilon^{2}_{V}(\varrho_{w})=\sum_{m,n=\pm}p_{m}p_{n}\text{
}S_{m}S_{n} \varrho_{w}(S_{m}S_{n})^{\dagger}=Tr_{c}\otimes
Tr_{c}(V\otimes V\varrho
_{c}\otimes\varrho_{c}\otimes\varrho_{w}V^{\dagger}\otimes
V^{\dagger}).
\end{equation}

This unitary extension of $\ \varepsilon^{2},$ requires a double
ancilla space or two coin quantum systems coupled, so the total
space is $H_{c}\otimes H_{c}\otimes H_{w}.$ In the general case
the $k$th power of the CPTP map of an ARW can be implemented
unitarily by extending the original \textit{walker space }by $k$
anchillary coin systems,\textit{ }$\ $so the total space becomes$\
H_{c}^{\otimes k}\otimes H_{w},$ and the total unitary
operator is%
\begin{equation}
V^{\otimes
k}\equiv\sum_{m_{1},...,m_{k}=\pm}P_{m_{1}}U\otimes...\otimes
P_{m_{k}}U\otimes S_{m_{1}}...S_{m_{k}}%
\end{equation}
with unitarity condition $V^{\otimes k}V^{\otimes k\dagger}=\mathbf{1}%
_{c}^{\otimes k}\otimes\mathbf{1}_{w}.$ Then we obtain for the
$k$th step of the QRW as described by $k$ successive actions of
its CPTP map, a unitary realization which involves tensoring of
\textit{quantum walker }to $k$ quantum \textit{coin systems},
followed by a coupling of them by a unitary evolution operator on
the space of coins+walker composite system, and finally a
decoupling of coins from the walker system, taken by partially
tracing with respect to the coin Hilbert spaces. The partial
tracing corresponds to coin tossing in an ordinary random walk,
and results into a density matrix for the quantum system of the
walker, which further may provide statistics of various quantum
observables of the walk.

Explicitly the unitarization of ARW reads%
\bea \varepsilon^{k}_{V}(\varrho_{w})&=&
\sum_{m_{1},...,m_{k}=\pm}p_{m_{1}}...p_{m_{k}
}S_{m_{1}}...S_{m_{k}}\varrho_{w}(S_{m_{1}}...S_{m_{k}})^{\dagger}\nn
\\  &=& Tr_{c}^{\otimes k}(V^{\otimes k}\varrho_{c}^{\otimes
k}\otimes\varrho _{w}V^{\otimes k\dagger}), \eea
and can be
identified with a QRW.

Let us remark at this point that an equivalent decomposition of
the
unitary $V^{\otimes k}$ would be%
\begin{equation}
V^{\otimes k}=\prod_{i=1}^{k}W_{i},\text{ where }W_{i}=(P_{+}U)_{i}%
\bigotimes(S_{+})_{k+1}+(P_{-}U)_{i}\bigotimes(S_{-})_{k+1},
\end{equation}
and the subindex denotes the position of the embedding of the
respective operator into the $k$-fold tensor product. In fact each
of these operators $W_{i}\in End(H_{c}^{\otimes k}\bigotimes
H_{w}),$ provide a new decomposition of the CPTP map i.e
$\varepsilon_{V}^{k}=\varepsilon_{W_{1}W_{2}}..._{W_{k}}$ , which
is equivalent to a nonstationary QRW with $k$ different unitary
evolution operators empoyed in order to construct the $k-th$ step.
As a matter of fact this new decomposition helps to account for
the type of quantum entanglement involved between coin and walker
systems. Let us take the simplest $k=2$ case,
where $V^{\otimes2}=W_{1}W_{2},$ with%
\begin{align}
W_{1} &  =\left(  P_{+}\bigotimes\mathbf{1}_{c}\bigotimes S_{+}+P_{-}%
\bigotimes\mathbf{1}_{c}\bigotimes S_{-}\right)  U\bigotimes\mathbf{1}%
_{c}\bigotimes\mathbf{1}_{w}\mathbf{,}\nonumber\\
W_{2} &  =\left(  \mathbf{1}_{c}\bigotimes P_{+}\bigotimes S_{+}%
+\mathbf{1}_{c}\bigotimes P_{-}\bigotimes S_{-}\right)  \mathbf{1}%
_{c}\bigotimes U\bigotimes\mathbf{1}_{w}\mathbf{.}%
\end{align}
The action of these operators on the product density matrices
$\rho_{c}\bigotimes\rho_{c}\bigotimes\rho_{w}$ is akin to
theaction of some control-control-$S_{\pm}$ type of non-local
operator, which uses the two coin states as \textit{control
spaces} and the walker state as the \textit{target space},
preceded by the local unitary operator $U$ which acts on the
control spaces and creates appropriate superposition of coin
states. These actions generate quantum entanglement and can be
described by the quantum circuit of Fig.1.c below. For purpose of
comparison in Fig.1.a, we have included the corresponding circuit
that generates the four entangled
bipartite Bell states upon action of the composite operator $U_{CN}%
H\bigotimes\mathbf{1}$ on the four orthogonal product qubit
states, and on Fig. 1b the circuit that corresponds to the unitary
$V=(P_{+}\bigotimes S_{+}+P_{-}\bigotimes
S_{-})U\bigotimes\mathbf{1,}$ of the $\varepsilon_{V}$ map.

\bigskip
\begin{figure}[h]
\begin{center}
\includegraphics[height=10cm, width=11cm]{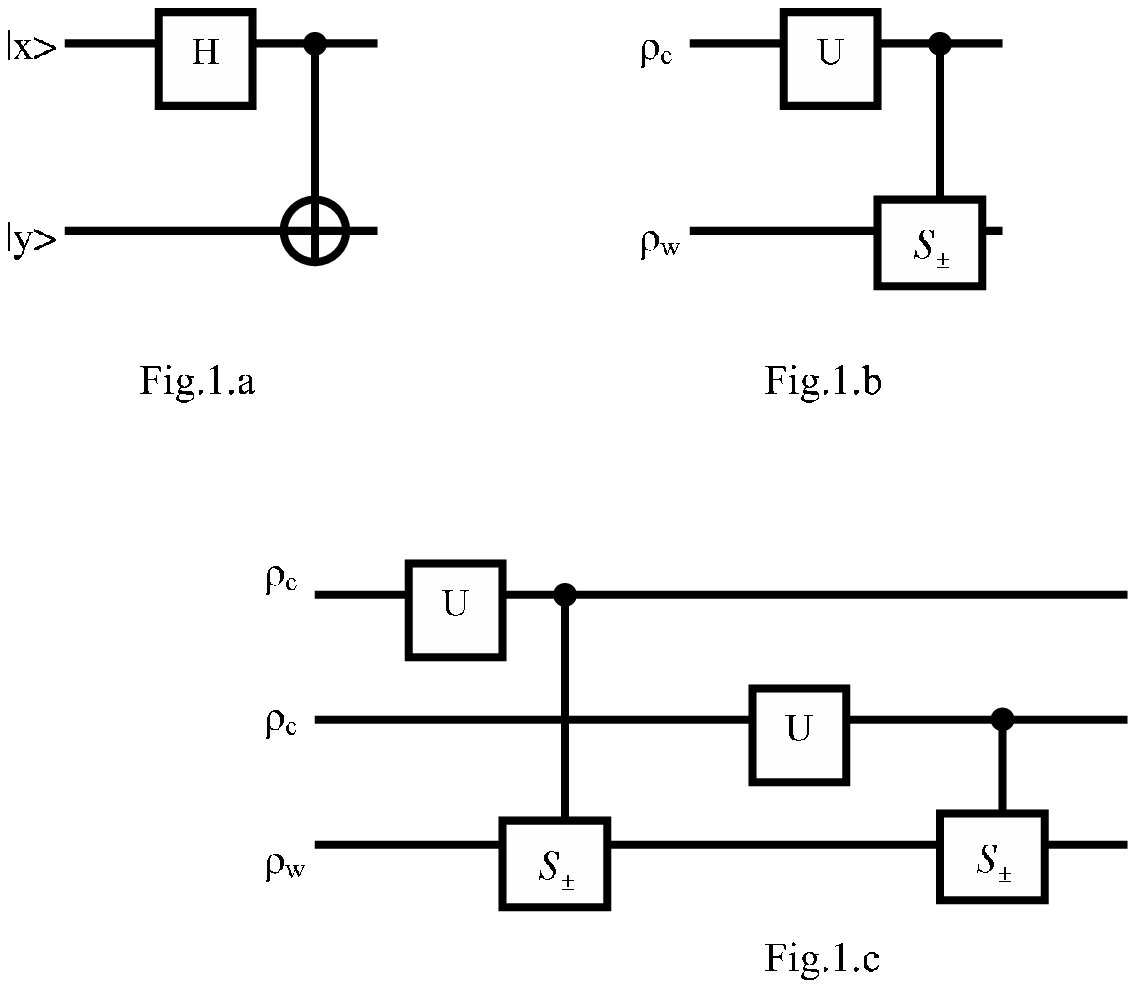}
\end{center}
\caption{ Fig.1 (a) Circuit generating entangled Bell states from
product states using local Hadamard and non-local control-not
gate; (b) Circuit generating the unitary evolution of the 1-step
map $\varepsilon_{V}$ ; (c) Circuit generating entangled
coin-walker states from corresponding factorized ones using local
unitary (e.g Hadamard gate in the case of the homonymous QRW), and
non-local control-control-$S_{\pm}$ gates, that form the map of
the 2-step QRW $\varepsilon_{V}^{2}=\varepsilon_{W_{1}W_{2}}$.}

\label{inter}
\end{figure}

Though the topic of the entanglement in the QRWs will not be
investigated further here, it should be obvious from the above
analysis that the CPTP map that implements some discrete time-step
of a QRW, also generates quantum entanglement that can be studied
by appropriate circuits and evaluated by effective measures, as
usually is done in other cases of coupled quantum systems.
Although quantum correlations have been generally accepted to be
the common cause of all novel effects in the \ \ QRW performance,
the exact evaluation of the entanglement resources needed in the
course of a QRW \ \ \ is still an open problem.

To establish further the connection among ARQs and QRWs we give
four particular examples:

\textit{i)} the Euclidean QRW \cite{bet} [\textbf{iso(2)}-QRW]:
with the \textit{distance operator} $L\left\vert m\right\rangle
=m\left\vert m\right\rangle ,$ with its eigenspace
$H_{w}^{L}=$span$\{\left\vert m\right\rangle ,m\in\mathbf{Z}\}$,
and its dual \textit{phase operator} $\Phi\left\vert
\varphi\right\rangle =\varphi\left\vert \varphi\right\rangle ,$
with its eigenspace $H_{w}^{\Phi}=$span$\{\left\vert
\varphi\right\rangle
,\varphi\in\lbrack0,2\pi);\frac{d\varphi}{2\pi}\},$ related by a
Fourier transform with $H_{w}^{L}.$ Two ARWs and its associated
QRWs (modulo local unitary operators in coin spaces, as explained
above), can be constructed: the \textit{distance }random walk on
$\mathbf{Z},$ with CPTP map constructed with Kraus generators been
the step operators in the distance operator eigenstates i.e $\
S_{\pm}\left\vert m\right\rangle \equiv E_{\pm}\left\vert
m\right\rangle =e^{\pm i\Phi}\left\vert m\right\rangle =\left\vert
m\pm1\right\rangle ,$ and the \textit{phase }random walk on the
circle $S,$ with CPTP map constructed with Kraus generator been
the step operators in the phase operator eigenstates i.e $\ $
$S_{\pm}\left\vert \varphi\right\rangle \equiv e^{\pm
iL}\left\vert \varphi\pm1\right\rangle ;$

\textit{ii)} the Canonical Algebra QRW [ \textbf{hw}-QRW]: with
the \textit{position operator} $Q\left\vert q\right\rangle
=q\left\vert q\right\rangle ,$ with its eigenspace
$H_{w}^{Q}=$span$\{\left\vert q\right\rangle
,q\in\mathbf{R};dq\},$ and its dual \textit{momentum operator}
$P\left\vert p\right\rangle =p\left\vert p\right\rangle ,$ with
its eigenspace $H_{w}^{P}=$span$\{\left\vert p\right\rangle
,p\in\mathbf{R};dp\},$ related by a Fourier transform with
$H_{w}^{Q}.$ Two ARWs and its associated QRWs (modulo local
unitary operators in coin spaces, as explained above), can be
constructed: the \textit{position }random walk on $\mathbf{R},$
with CPTP map constructed with Kraus generators been the step
operators in the position operator eigenstates i.e $\ $
$S_{\pm}\left\vert q\right\rangle \equiv e^{\pm iP}\left\vert
q\right\rangle =\left\vert q\pm1\right\rangle ,$ and the
\textit{momentum }random walk on $\mathbf{R},$ with CPTP map
constructed with Kraus generator been the step operators in the
momentum operator eigenstates i.e $\ $ $S_{\pm}\left\vert
p\right\rangle \equiv e^{\pm iQ}\left\vert p\right\rangle
=\left\vert p\pm1\right\rangle ;$

\textit{iii)} the $M-$ dimensional Discrete Heisenberg Group QRW
[\textbf{h}$_{M} $-QRW]: with the \textit{action operator}
$N\left\vert n\right\rangle =n\left\vert n\right\rangle ,$ with
its eigenspace $H_{w}^{N} =$span$\{|n>,n\in\mathbf{Z}_{M}\},$ and
its dual \textit{angle operator} $\Theta\left\vert
\vartheta_{n}\right\rangle =\vartheta_{n}\left\vert
\vartheta_{n}\right\rangle ,$ with its eigenspace $H_{w}^{\Theta}
=$span$\{\left\vert \vartheta_{n}\right\rangle ,$
$\vartheta_{n}\in\frac {1}{2\pi}\mathbf{Z}_{M}\},$ related by a
finite Fourier transform with $H_{w}^{N}.$ Two ARWs and its
associated QRWs (modulo local unitary operators in coin spaces, as
explained above), can be constructed: the \textit{action }random
walk on $\mathbf{Z}_{M},$ with CPTP map constructed with Kraus
generators been the step operators in the action operator
eigenstates i.e $\ $ $\ S_{\pm}\left\vert n\right\rangle \equiv
e^{\pm i\Theta}\left\vert n\right\rangle =h^{\pm1}\left\vert
n\right\rangle =\left\vert n\pm 1\right\rangle ,$ and the
\textit{angle }random walk on $\frac{1}{2\pi }\mathbf{Z}_{M},$
with CPTP map constructed with Kraus generator been the step
operators in the momentum operator eigenstates i.e $\ $
$S_{\pm}\left\vert \vartheta_{n}\right\rangle \equiv
g^{\pm1}\left\vert \vartheta_{n} \right\rangle =e^{\pm\not i
N}\left\vert \vartheta_{n}\right\rangle =\left\vert
\vartheta_{n\pm1}\right\rangle ;$

\textit{iv)} the Coherent State QRW \cite{appell}[
\textbf{CS}-QRW]: with the \textit{annihilation operator}
$a\left\vert \alpha\right\rangle =\alpha\left\vert
\alpha\right\rangle ,$ with its eigenspace $H_{w}^{a}
=$span$\{\left\vert \alpha\right\rangle ,\alpha\in\mathbf{C};\frac{d^{2}%
\alpha}{2\pi}\}.$Two ARWs and its associated QRWs (modulo local
unitary operators in coin spaces, as explained above), can be
constructed: the \textit{annihilation }random walk on
$\mathbf{C},$ with CPTP map constructed with Kraus generators been
the step operators in the annihilation operator eigenstates i.e $\
S_{\pm}\left\vert \alpha\right\rangle \equiv e^{J(\pm
1)}\left\vert \alpha\right\rangle =D_{\pm1}\left\vert
\alpha\right\rangle =\left\vert \alpha\pm1\right\rangle .$ The
step operators here identified as special case of the canonical
coherent state displacement operator $D_{\pm\beta}=e^{\pm\beta
a^{\dagger}\mp\overline{\beta}a}=e^{J(\pm\beta)},$ have based the
indicated step property on the following operator identity
$D_{\alpha}D_{\beta}=e^{-i\alpha\times\beta}D_{\alpha+\beta},$
applied for the case of co-linear $\alpha,\beta$ vector on complex
plane.

To give an explicit identification of the ARW based on the
\textbf{hw} algebra constructed in \cite{appell}, as a quantum
random walk, and in particular as a \textbf{CS}-QRW, we make the
following choices: the transition probabilities are
$p_{+}=p,p_{-}=1-p$, the coin state is $\rho_{c}=|0><0|,$ the $U$
operator is $U(p,1-p)$, and the step operators are CS displacement
operators with steps $\pm\beta\in\mathbf{C}.$ Then the total
1-step evolution operator in the coin+walker system is
$V=\sum_{m=\pm}P_{m}U\otimes S_{m}=\left(
\begin{array}
[c]{cc}
\sqrt{p}D_{+\beta} & \sqrt{1-p}D_{+\beta}\\
\sqrt{1-p}D_{-\beta} & -\sqrt{p}D_{-\beta}
\end{array}
\right)  ,$ and the reduced walker evolves in 1-step by the CPTP
$\varepsilon_{V}(\varrho_{w})=pD_{+\beta}\varrho_{w}D_{+\beta}^{\dagger
}+(1-p)D_{-\beta}\varrho_{w}D_{-\beta}^{\dagger}=Tr_{c}(V\varrho_{c}
\otimes\varrho_{w}V^{\dagger}).$

For $n$ steps the evolution of the walker has been chosen in
\cite{appell} to be $\varepsilon_{V}^{n}(\varrho_{w}).$ It is
important to notice that this is a choice based on the ARW
construction methodology, and that our present treatment of the
same walk as a QRW, sees the $\varepsilon_{V}^{n}(\varrho _{w})$
type of evolution to result from a partial tracing of the coin
system at every step. Our previous discussion of other types of
tracing schemes motivates the study of \textbf{CS}-QRWs with
delayed tracing, in order to investigate phenomena such as
enhanced or anomalous diffusion in ARWs. This problem will be
taken up elsewhere.

\section{Discussion}
We have outlined a mathematical framework where the conception of
random walk and its associated statistical notions, and equations
of motion, can both be studied in an algebraic and quantum
mechanical manner. ARWs and QRWs appear to be two aspects of the
same mathematical device, so their interconnection serves to
conceptually clarify the common ground between them and to enrich
the heuristics of formulating new problems and methodically
searching for their solutions.

Quantum random walks are important both as quantum algorithms to
experimentally be realized and as modules in a general quantum
computing algorithm-devise that could outperform some classical
rival. The connection ARW-QRW could serve to generalize, unify and
compare such algorithms.

Also Quantum Information Processing concepts and tools, could be
developed for ARW-QRWs. The step taken here is only a preliminary
one towards developing such a theory.

Finally, ARW-QRWs come with lots of free choices for its
constituting parameters. To mention only one expected application
in the field of Open Quantum Systems, we should emphasize the
importance of choosing the functional in e.g the \textbf{hw} ARW.
Various choices of functionals in terms of  types of coherent
state vectors, combined together with various choices of ordering
the operator basis in the enveloping algebra
$\mathcal{U}(\mathbf{hw})$, i.e normal, antinormal, symmetric etc,
could serve as a guiding rule for constructing quantum master
equations for open boson systems interacting with various types of
quantum mechanical baths.

\section{Acknowledgments} I wish to thank the organizers of
the Volterra-CIRM-Grefswald Conference for the opportunity to give
a talk. Discussions with L. Accardi, U. Franz, R. Hudson, M.
Sch\"urmann, and with my collaborators A. Bracken and I.
Tsohantjis, are gratefully acknowledged. I am also grateful  to
the anonymous referee for suggesting eq.(63).

\pagebreak


\begin{thebibliography}{99}                                                                                               %

\bibitem{meyer}
P. A. Meyer, {\it Quantum Probability for Probabilists} (Lect.
Notes Math. 1538), (Springer, Berlin 1993).
\bibitem{schurmann}
M. Sch\"urmann, {\it White Noise on Bialgebras} (Lect. Notes Math.
1544), (Springer, Berlin 1993).
\bibitem{majidbook}
S. Majid, {\it Foundations of Quantum Groups Theory} (Cambridge
Univ. Press, 1955), ff. chapter 5.
\bibitem{fs}
U. Franz and R. Schott, \textit{Stochastic Processes
and Operator Calculus on Quantum Groups}, (Kluwer Academic
Publishers, Dodrecht 1999).

\bibitem{ambainis}
A. Ambainis, E. Bach, A. Nayak, A. Vishwanath and J. Watrous,
\textit{Proc. 33rd Annual Symp. Theory Computing } (ACM Press, New
York, 2001), p.37.

\bibitem{daharonov}
D. Aharonov, A. Ambainis, J. Kempe and U. Vasirani,
\textit{Proc. 33rd Annual Symp. Theory Computing } (ACM Press, New
York, 2001), p.50.

\bibitem{nayak}
A. Nayak and A. Vishwanath, arXive eprint quant-ph/0010117.

\bibitem{kempe}
J. Kempe, {\em Proc. 7th Int. Workshop, RANDOM'03}, p.354 (2003).

\bibitem{childs1}
A. M. Childs, E. Farhi and S. Gutmann,
\textit{Quantum Information Processing} \textbf{1}, 35 (2002).

\bibitem{travaglione}
 B. C. Travaglione and G. J. Milburn, \textit{Phys. Rev. A}
\textbf{65}, 032310 (2002).

\bibitem{sanders}
B. C. Sanders, S. D. Bartlett, B. Tregenna and P. L. Knight,
\textit{Phys. Rev. A} \textbf{67}, 042305(2003).

\bibitem{kempeReview}
J. Kempe, \textit{Contemp. Phys.} \textbf{44}, 307 (2003).

\bibitem{bet}
A. J. Bracken, D. Ellinas and I. Tsohantjis, \textit{J. Phys. A:
Math. Gen.} \textbf{37}, L91(2004).

\bibitem{abe} E. Abe, {\it Hopf Algebras} (CUP Cambridge 1997).

\bibitem{appell}D. Ellinas, \textit{J. Comp. Appl.
Math.}\textbf{133}, 341 (2001).

\bibitem{marshalolkin}  A. W. Marshall and I. Olkin, \textit{Inequalities:
Theory of Majorization and its Applications }(Academic Press, New
York, 1979).

\bibitem{bhatia}  R. Bhatia, \textit{Matrix Analysis} (Spinger-Verlag, New
York , 1997).

\bibitem{albertiuhlmann}  P. M. Alberti and A. Uhlmann, \textit{Stochasticity and
Partial Order: Double Stochastic Maps and Unitary Mixing}
(Dordecht, Boston, 1982).

\bibitem{nielsennotes}  M. A. Nielsen, \textit{An Introduction to Majorization
and its Applications to Quantum Mechanics} (unpublished notes).


\bibitem {ellinasf}
D. Ellinas and E. Floratos, \textit{J. Phys. A: Math. Gen.
}\textbf{32}, L63 (1999).


\bibitem {hughes}
B. D. Hughes, \textit{Random Walks and Random Environments
Vol. I, }(Clarendon Press, Oxford 1995)

\bibitem {gillis}
J. Gillis, \textit{Quarterly J. Math., (Oxford, 2nd
series)\textbf{7}}, 144 (1956).\textit{ }

\bibitem {ls}
K. Lakatos-Lindenberg and K. E. Shuler, \textit{J. Math. Phys.
}\textbf{12}, 633 (1971).

\bibitem{klauder}
J. R. Klauder and B.-S. Skagerstam, {\it Coherent States} (World
Scientific, Singapore (1986)

\bibitem{perelomov}
A. Perelomov, {\it Generalized Coherent States and their
Applications}, (Springer - Verlag, Berlin 1986).

\bibitem{davies}
E. D. Davies, {\it Quantum Theory of Open System}, (Academic, New
York, 1973).

\bibitem{lind}
G. Lindblad, {\it Non-Equilibrium Entropy and Irreversibility},
(Reidel, Dordrecht 1983).

\bibitem {scully}M. O. Scully and \ M. S. Zubairy, \textit{Quantum Optics,
}(Cambridge Univ. Press, Cambridge 1997), p. 255, 448, 453.

\bibitem {stenholm} S. Stenholm, \textit{Phys. Scripta} \textbf{T12}(1986).

\bibitem {gea} J. Gea-Banacloche, \textit{Phys. Rev.
Lett.} \textbf{59}, 543 (1987).

\bibitem{kraus}
K. Kraus, \textit{States, effects and operations}
(Springer-Verlag, Berlin, 1983).

\bibitem {nielsen}
M. A. Nielsen and I. L. Chuang, \textit{Quantum Computation
and Quantum Information}, (Cambridge Univ. Press, Cambridge 2000).

\bibitem{majid1}
S. Majid,  \textit{Int. J. Mod. Phys.} ${\bf 8}$ , 4521-4545
(1993).

\bibitem{majid2}
S. Majid, M. J. Rodriguez-Plaza, \textit{J. Math. Phys.} ${\bf
33}$, 3753-3760 (1994).


\bibitem{qnoice1}
P. Feinsilver and  R. Schott, \textit{J. Theor. Prob.} {\bf 5},
251 (1992).

\bibitem{qnoice2}P. Feinsilver, U. Franz and  R.
Schott, \textit{J. Theor. Prob.} {\bf 10}, 797 (1997).

\bibitem{qnoice3}U. Franz and R.
Schott, \textit{J. Phys. A: Gen . Math. }{\bf 31} , 1395 (1998);

\bibitem{qnoice4}U. Franz
and R. Schott, \textit{J. Math. Phys.} {\bf 39}, 2748 (1998).


\bibitem{elltso1}
D. Ellinas and I. Tsohantjis,  \textit{J. Non-Lin. Math.
Phys.}\textbf{8}Suppl. 93(2001).

\bibitem{elltso2}
D. Ellinas and I. Tsohantjis,\textit{Inf. Dim. Anal.- Quant.
Prob.} \textbf{6}11 (2003).

\bibitem {hardy}G. H. Hardy, J. E. Littlewood and G. Polya, \textit{Messenger
Math. }\textbf{58}, 145 (1929).

\bibitem {sikic}H. Sikic and M. V. Wickerhauser,
\textit{Appl. Comp. Harm. Anal.} \textbf{11}, 147 (2001).

\bibitem{gilmore}
R. Gilmore, {\it Lie Groups, Lie Algebras and Some of Their
Applications}, (Wiley, New York 1974).

\bibitem{ellinasetal}
D. Ellinas, et. al, work in progress.

\bibitem{circ}
P. J. Davis, {\it Circulant Matrices}, (Wiley, New York 1979).

\bibitem{schroeder}
M. R. Schroeder, {\it Number Theory in Science and Communication},
(Springer, Berlin 1997).

\bibitem{hudson}
C. D. Cushen and R. L. Hudson, \textit{J. Appl. Prob.} \textbf{8},
454 (1972).

\bibitem{comment}
This second line of the equation has the form a Lindblad type
master equation taken in the large temperature limit, where for
the average number of thermal photons we have
$\overline{n}\approx\overline{n}+1 = |\gamma|$, c.f.\cite{scully}.

\bibitem{risken}
H. Risken, {\it The Fokker-Planck Equation}, (Springer, Berlin
1996). 83 (1990).

\end{thebibliography}
\end{document}